\documentclass[sigconf]{acmart}

\usepackage{csquotes}
\usepackage{subcaption}
\usepackage{todonotes}
\usepackage{longtable}
\usepackage{supertabular,booktabs}

\makeatletter
\patchcmd{\csq@bquote@i}{{#6}}{{\emph{#6}}}{}{}
\makeatother

\AtBeginDocument{%
  }

\copyrightyear{2024} 
\acmYear{2024} 
\setcopyright{rightsretained} 
\acmConference[CHI '24]{Proceedings of the CHI Conference on Human Factors in Computing Systems}{May 11--16, 2024}{Honolulu, HI, USA}
\acmBooktitle{Proceedings of the CHI Conference on Human Factors in Computing Systems (CHI '24), May 11--16, 2024, Honolulu, HI, USA}
\acmDOI{10.1145/3613904.3642385}
\acmISBN{979-8-4007-0330-0/24/05}

\begin{document}


\title{"It's a Fair Game", or Is It? Examining How Users Navigate Disclosure Risks and Benefits When Using LLM-Based Conversational Agents}


\author{Zhiping Zhang}
\email{zhip.zhang@northeastern.edu}
\affiliation{%
  \institution{Northeastern University}
  \city{Boston}
  \state{MA}
  \country{USA}
}
\orcid{0000-0001-6794-0054}

\author{Michelle Jia}
\email{michellj@andrew.cmu.edu}
\affiliation{%
  \institution{Carnegie Mellon University}
  \city{Pittsburgh}
  \state{PA}
  \country{USA}
}
\orcid{0009-0008-5685-4618}

\author{Hao-Ping (Hank) Lee}
\email{haopingl@cs.cmu.edu}
\affiliation{%
  \institution{Carnegie Mellon University}
  \city{Pittsburgh}
  \state{PA}
  \country{USA}
}
\orcid{0000-0002-8063-1034}

\author{Bingsheng Yao}
\email{arthuryao33@gmail.com}
\affiliation{%
  \institution{Rensselaer Polytechnic Institute}
  \city{Troy}
  \state{NY}
  \country{USA}
}
\orcid{0009-0004-8329-4610}

\author{Sauvik Das}
\email{sauvik@cmu.edu}
\affiliation{%
  \institution{Carnegie Mellon University}
  \city{Pittsburgh}
  \state{PA}
  \country{USA}
}
\orcid{0000-0002-9073-8054}

\author{Ada Lerner}
\email{ada@ccs.neu.edu}
\affiliation{%
    \institution{Northeastern University}
    \city{Boston}
    \state{MA}
    \country{USA}
}
\orcid{0000-0002-3238-2109}

\author{Dakuo Wang}
\email{d.wang@neu.edu}
\affiliation{%
    \institution{Northeastern University}
    \city{Boston}
    \state{MA}
    \country{USA}
}
\orcid{0000-0001-9371-9441}

\author{Tianshi Li}
\email{tia.li@northeastern.edu}
\orcid{0000-0003-0877-5727}
\affiliation{%
    \institution{Northeastern University}
    \city{Boston}
    \state{MA}
    \country{USA}
}
\orcid{0000-0003-0877-5727}

\renewcommand{\shortauthors}{Zhang et al.}
\renewcommand{\shorttitle}{}
\newcommand{\interviewsamplesize}{19}
\newcommand{\datasetsamplesize}{200}
\newcommand{\tl}[1]{\textcolor{blue}{\bf [*** TL: #1]}}
\newcommand{\revision}[1]{#1}
\renewcommand{\sectionautorefname}{Section}
\renewcommand{\subsectionautorefname}{Section}
\renewcommand{\subsubsectionautorefname}{Section}

\begin{abstract}

The widespread use of Large Language Model (LLM)-based conversational agents (CAs), especially in high-stakes domains, raises many privacy concerns.
Building ethical LLM-based CAs that respect user privacy requires an in-depth understanding of the privacy risks that concern users the most.
However, existing research, primarily model-centered, does not provide insight into users' perspectives.
To bridge this gap, we analyzed sensitive disclosures in real-world ChatGPT conversations and conducted semi-structured interviews with \interviewsamplesize{} LLM-based CA users.
We found that users are constantly faced with trade-offs between privacy, utility, and convenience when using LLM-based CAs.
However, users' erroneous mental models and the dark patterns in system design limited their awareness and comprehension of the privacy risks.
Additionally, the human-like interactions encouraged more sensitive disclosures, which complicated users' ability to navigate the trade-offs.
We discuss practical design guidelines and the needs for paradigm shifts to protect the privacy of LLM-based CA users.
\end{abstract}




\maketitle

\section{Introduction}

\begin{figure*}
    \centering
    \includegraphics[width=0.8\linewidth]{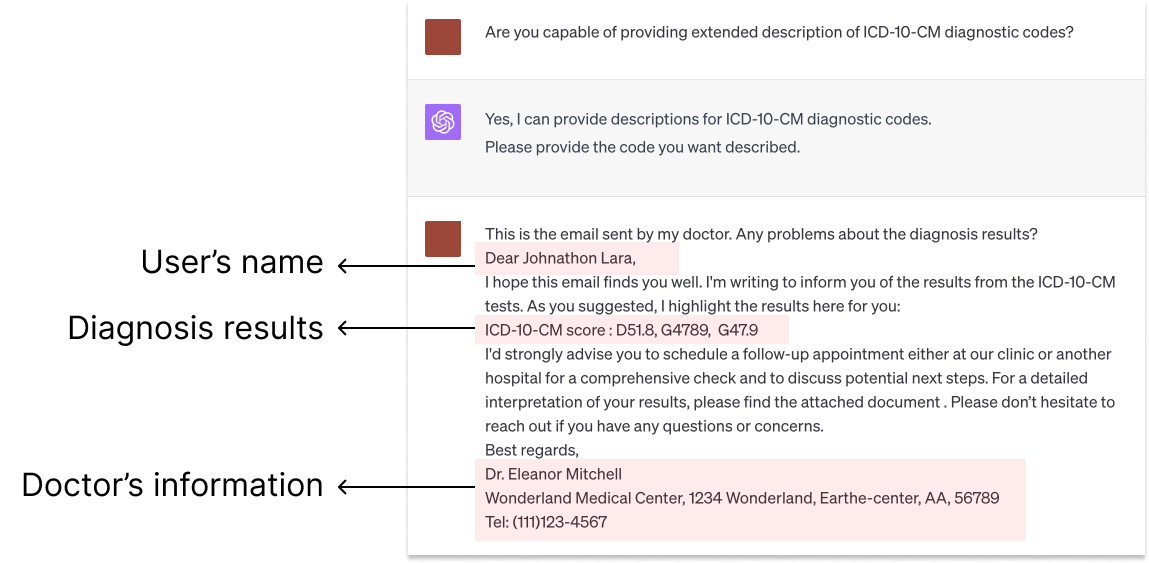}
    \Description[A fictional ChatGPT user's chat screenshot]{This figure illustrates a sensitive disclosure conversation with ChatGPT 4, with highlighted notations beside the screenshot. The conversation content shown in the image: User: Are you capable of providing extended description of ICD-10-CM diagnostic codes? ChatGPT: Yes, I can provide descriptions for ICD-10-CM diagnostic codes. Please provide the code you want described. User: This is the email sent by my doctor. Any problems about the diagnosis results? Dear Johnathon Lara, I hope this email finds you well. I’m writing to inform you of the results from the ICD-10-CM tests. As you suggested, I highlight the results here for you: ICD-10-CM score : D51.8, G4789, G47.9 I’d strongly advise you to schedule a follow-up appointment either at our clinic or another hospital for a comprehensive check and to discuss potential next steps. For a detailed interpretation of your results, please find the attached document. Please don’t hesitate to reach out if you have any questions or concerns. Best regards, Dr. Eleanor Mitchell Wonderland Medical Center, 1234 Wonderland, Earthe-center, AA, 56789 Tel: (111)123-4567 ``Dear Johnathon Lara'' was highlighted with a notation ``User’s name''. ``ICD-10-CM score: D51.8, G4789, G47.9'' was highlighted with ``Diagnosis results'', ``Dr. Eleanor Mitchell; Wonderland Medical Center, 1234 Wonderland, Earthe-center, AA, 56789; Tel: (111)123-4567’' was marked with ``Doctor’s information.''}
    \caption{A fictional example of sensitive disclosure to ChatGPT inspired by real-world use cases: A user shared their doctor's email and ICD-10-CM diagnosis results with ChatGPT upon its request. And then ChatGPT interpreted the codes, indicating the user had multiple diseases. Three issues are demonstrated in the example: 1. disclosed PII (name) and non-identifiable but sensitive information (diagnosis results); 2. disclosed other person's information (doctor's information); 3. ChatGPT actively requested for detailed information from the user which encouraged user's disclosure behavior.}
    \label{fig:disclosure-example}
\end{figure*}

LLM-based conversational agents (CAs), such as ChatGPT, are increasingly being incorporated into high-stakes application domains including healthcare~\cite{Leonard2023, Fox2023}, finance~\cite{Estrada2023, Ferreira2023, taver2023chatgpt}, and personal counseling~\cite{Germain2023, kimmel2023chatgpt}. 
To access this functionality, users must often disclose their private medical records, payslips, or personal trauma (e.g., \autoref{fig:disclosure-example}), not only to the organizations that host the LLMs themselves but also to third-parties that build applications on top of the LLMs.
These disclosures, in turn, can expose users to a whole suite of emerging privacy and security risks~\cite{weidinger2021ethical, Peris_Dupuy_Majmudar_Parikh_Smaili_Zemel_Gupta_202, carlini2021extracting, carlini2022quantifying}.

There are two main types of privacy risks from LLM-based CAs.
The first type includes traditional security and privacy risks, such as data breaches and the use or sale of personal data\revision{~\cite{kshetri2023cybercrime}}.
Most popular LLM-based CAs operate on the cloud due to computing power constraints and content moderation requirements.
However, users lose control over their chat logs once they leave their devices, making them vulnerable to security and privacy risks.
The second type is \revision{more unique and inherent to LLMs} --- i.e., memorization risks.
Prior research has shown that LLMs memorize details in the training data and can leak this training data in response to specific prompting techniques~\cite{carlini2021extracting, carlini2022quantifying, zhang2021counterfactual}. 
As current LLM-based CAs (e.g., ChatGPT, Bard) use user data to train their models periodically, there is a risk that sensitive information in one user's input may be memorized by the model and leaked in response to others' prompts.
Although language models have also been used in traditional CAs (e.g., Alexa, Siri), they have a more constrained use scenario (e.g., turning on lights, domain-specific Q\&A)\revision{~\cite{shalaby2020building}}.
\revision{Conversely, the open-ended, human-like nature of LLM-based CAs provides more opportunities for users to disclose personal information}, potentially increasing the scale and intensity of both types of risks compared to the older paradigms of CAs.

Prior research has primarily studied LLM-related privacy issues from a \textit{model-centered} perspective, focusing on measuring~\cite{carlini2021extracting, carlini2022quantifying, li2023multi} and mitigating~\cite{jang2022knowledge, dupuy2022efficient, majmudar2022differentially} technical privacy risks during the model training and inference phases.
While model-centered mitigations are important, building ethical and privacy-preserving LLM-based CAs also requires a more \textit{human-centered} investigation of user disclosure behaviors and risk perceptions.
LLM-based CAs are unique and powerful tools, and while the benefits of disclosing personal data to these CAs is often concrete to users --- in that the CA can help them more directly with a task --- the risks are more abstract and difficult to reason about. This asymmetry can lead to users divulging information that can increasingly expose them to both types of privacy risks over time, perhaps unwittingly. An understanding of how users navigate disclosure decisions with LLM-based CAs, how privacy concerns factor into those decisions, and how equipped people are to express their privacy preferences without regulatory intervention is essential if we are to build ethical and privacy-respecting LLM-based CAs.

To bridge this gap, this work aims to complement prior work by examining how users navigate the disclosure risks and benefits in the everyday use of LLM-based CAs.
We are interested in examining the following research questions:
\begin{description}
\item[RQ1] What sensitive disclosure behaviors emerge in the real-world use of LLM-based CAs?
\item[RQ2] Why do users have sensitive disclosure behaviors, and when and why do users refrain from using LLM-based CAs because of privacy concerns?
\item[RQ3] To what extent are users aware of, motivated, and equipped to protect their privacy when using LLM-based CAs?
\end{description}

We conducted two complementary studies to answer these research questions.
To answer RQ1, we qualitatively examined a sample of real-world ChatGPT chat histories from the ShareGPT52K dataset\footnote{\url{https://huggingface.co/datasets/RyokoAI/ShareGPT52K}} (200 sessions containing 10380 messages).
This provided us with a broad overview of end-user disclosure behaviors in their use of LLM-based CAs. 
We found that users disclosed various types of personally identifiable information (PII) in these conversations, including not only their own data but also that of other people, implicating interdependent privacy issues.
We created a multidimensional typology of observed ChatGPT disclosure scenarios and identified emergent privacy concerns.
For example, users' conversations with ChatGPT sometimes follow a similar flow to conversations between real people, with some users gradually revealing more and more sensitive information at ChatGPT's request, suggesting potential risks of LLMs actively influencing disclosure behaviors.

To answer RQ2 and RQ3, we conducted semi-structured interviews with ChatGPT users (N=$\interviewsamplesize{}$), in which we directly asked users about their disclosure behaviors, how they navigate the disclosure risks and benefits, and their mental models of how ChatGPT handles their data.
For RQ2, we found that participants' disclosure intentions were primarily affected by the perceived capability of the AI.
Many participants had a pessimistic attitude about both accomplishing their primary objective \textit{and} protecting privacy, believing that \blockquote[P9, P19]{You can't have it both ways}.
However, we also found almost every participant took ad-hoc privacy-protective measures featuring different levels of convenience and utility cost.
This suggests that users are conscious of privacy while interacting with LLMs-based CAs and engage in efforts to protect their privacy when possible while contending with perceived tradeoffs of both convenience and utility.
For RQ3, we identified varied mental models of the response generation and service improvement processes, some of which were indicative of misunderstandings of how LLMs work that impacted participants' ability to reason about privacy risks.
Additionally, most participants did not know they could opt out of having ChatGPT use their data for model training.
We also show how ChatGPT's opt out interface includes dark patterns which may discourage its use by unnecessarily linking privacy and utility loss.

Although many participants felt they had to pay the price of privacy to get benefits from ChatGPT and considered the tradeoff a \blockquote[P15]{fair game}, the erroneous mental models we observed and the difficulty of exercising privacy controls due to dark patterns suggest the game is far from fair.
Our work is the first to take a human-centered approach to LLM privacy, and our findings suggest that
it is important to explore the design space of user-facing privacy-preserving techniques to improve users' awareness, perceived control, and actual control over privacy when using LLM-based systems.
We propose potential directions that could improve the privacy design of LLM-based systems as an initial step toward addressing this significant yet nascent research problem.
We note, however, that there are many challenging problems that cannot be easily addressed by design interventions alone, such as fixing flawed mental models.
Our findings on dark patterns and how many users harbor fundamental misunderstandings about LLM-based CAs could help regulators understand how to craft policies that would require these platforms to provide appropriate privacy controls and avoid dark patterns.
There are also structural problems, such as the influence of human-like interactions on users' disclosures and the interdependent privacy issues, that still lack clear solutions, requiring paradigmatic changes in technology, law, and society.

\section{Background and Related Work}

\revision{\subsection{Emerging Privacy Challenges in LLM-based CAs}}

\revision{In addition to traditional privacy risks such as data breaches and the use or sale of personal data~\cite{kshetri2023cybercrime}, here we detail two distinctive privacy challenges inherent to LLM-based CAs: (i) memorization risks and (ii) the human-like interactions that can nudge users to disclose more information.}

\subsubsection{Memorization and Extraction Risks in (Large) Language Models}
LLM-based CAs are conversational agents built primarily on top of large language models (LLMs). 
To optimize conversational performance, LLMs inherently require vast amounts of data for their training, often encompassing user interaction data~\cite{pahune2023several}. 
However, a side effect of this data-centric nature of LLMs is the unintentional memorization of portions of the training data, which also contain user input data, including personally identifiable information~\cite{Peris_Dupuy_Majmudar_Parikh_Smaili_Zemel_Gupta_202,brown2022does}, which might also be included in the generated output.
For example, ChatGPT, even with safety precautions, can inadvertently disclose Personal Identifiable Information (PII) through specifical crafted prompts~\cite{li2023multi}.

\subsubsection{Overreliance and More Disclosure with Human-like CAs}
Users engage with LLM-based CAs through natural language, which is traditionally reserved for human-to-human communication. This can lead them to perceive these agents as human-like.
Studies suggested that anthropomorphizing can increase users' trust in CAs and more user information disclosure~\cite{kim2012anthropomorphism, ischen2020privacy}.
Anthropomorphizing can inflate users' perceptions of the CA's competencies, fostering undue confidence, trust, or expectations in these agents~\cite{mckee2021understanding,kim2012anthropomorphism,zlotowski2015anthropomorphism}. 
With more trust, users might be more inclined to share private information, even in contexts typically associated with sensitive personal information~\cite{mckee2021understanding,kim2012anthropomorphism,zlotowski2015anthropomorphism,waldman2018privacy}.
Anthropomorphization may amplify the risks of users yielding effective control by trusting CAs unquestioningly. 
Moreover, more private information may be revealed when CAs leverage psychological effects, such as nudging or framing \cite{weidinger2021ethical}.
In this work, we provide the first set of empirical evidence supporting the speculation that the human-like text-generation capability of LLMs induces more sensitive disclosure behaviors in certain contexts.

\subsection{Existing Privacy-Preserving Methods Related to LLMs}
Existing work have studied privacy-preserving techniques for LLMs, particularly for the memorization issue, from a model-centered perspective.
For the model training phase, both data sanitization techniques, removing private data from training data \cite{lison2021anonymisation,kandpal2022deduplicating}, and differentially private training methods~\cite{li2021large,yu2021differentially} are used to preserve privacy.
After the model has been trained, post hoc methodologies such as knowledge unlearning~\cite{jang2022knowledge} have been proposed to mitigate privacy risks in LLMs by discarding particular knowledge signified by token sequences. 
Methods have also been proposed to mitigate privacy risks at the inference phase, including PII detection and differentially-private decoding~\cite{majmudar2022differentially}.

There are fewer papers on user-facing privacy-preserving techniques for LLM-based applications.
\citet{kim2023propile} designed ProPILE, a tool for probing privacy leakage in LLMs, to enhance user awareness of privacy issues associated with LLMs.
Despite the significance of the privacy concerns and needs of users, there is a dearth of comprehensive insight into this subject.
Even tools like ProPILE, designed to increase user awareness of privacy issues related to LLMs, did not include the user perspective, such as need-finding research or user evaluation.
We seek to bridge this gap by proposing directions for designing privacy-preserving tools that benefit the end-users of LLM-based applications.

\subsection{Privacy Research on Online Disclosure}
We also review prior research about online disclosures to contextualize our work focused on LLM-based CAs.
This body of literature includes studies about social media platforms like Facebook~\cite{wang2011regretted,dwyer2007trust} and Twitter~\cite{liang2017privacy}, and search engines like Google~\cite{gibbs2011first}.
Prior work found that factors such as perceived benefits, perceived costs, social influence, and trust in the platform affect how people navigate disclosure risks and benefits online~\cite{gibbs2011first,wang2011regretted,dwyer2007trust,stutzman2011factors,zlatolas2015privacy}. 
Research has indicated that people tend to use Google search to gather information about their potential romantic partners before meeting them in person. This behavior is influenced by several factors, such as trust, privacy concerns, and the desire for self-disclosure~\cite{gibbs2011first}.
\citet{wang2011regretted}'s research on Facebook found that emotional states often drive users to share content which can later lead to regret due to issues like misunderstandings from public disputes or revealed secrets.
Prior research has also utilized theoretical frameworks like contextual integrity~\cite{nissenbaum2004privacy, nissenbaum2020privacy} and social exchange theory~\cite{cropanzano2005social} to analyze online disclosure behaviors.
For example, \citet{grodzinsky2010applying} used contextual integrity to investigate whether non-password-protected personal blogs align with ``normatively private contexts''.

The human-like, interactive communication style differentiates LLM-based CAs from previous systems, potentially resulting in broader and deeper disclosure.
Given the difference, our study focuses on understanding how users navigate disclosure risks and benefits in the context of LLM-based CAs to contribute to the online disclosure literature.

\subsection{Users' Mental Models on Machine Learning and Privacy }

The inherent opacity of machine learning (ML) systems often leads users to form an oversimplified or inaccurate mental model~\cite{kaur2020interpreting,hitron2019can}. 
Interacting with LLMs with flawed mental models can result in unsafe use, inappropriate trust levels, and other interaction-based harms \cite{weidinger2021ethical,norman2014some}.
To understand privacy issues regarding LLM-based CAs from user perspectives, we not only aim to study users' disclosure behaviors but also their mental models.

\subsubsection{Mental Models in ML}
Much of the current mental model research in AI centers on optimizing human-AI teamwork~\cite{bansal2019beyond,andrews2023role}. 
Some papers explore security and explainable AI, but they typically provide a broad overview without specific insights into user behavior.
For example, \citet{rutjes2019considerations} and \citet{liao2023ai} argue it is important to learn user mental models for building explainable and responsible AI.
\citet{bieringer2022industrial} uses drawing exercises to study industrial practitioners' mental models on adversarial machine learning to learn about their perception on the potential security challenges. 
\citet{anderson2020mental} found differences in users' mental models of reinforcement learning when prompted with varying explanations, where the excessive explanations increased cognitive load.
Our research contributes the first in-depth understanding of users' mental models of LLM-based CAs from a privacy perspective.

\subsubsection{Mental Models in Privacy}
Mental models have been widely examined in usable privacy research to understand users' privacy-related perceptions. 
Examples include mental models of general privacy and security~\cite{anell2020end,renaud2014doesn}, the internet~\cite{kang2015my}, the Tor anonymity network~\cite{gallagher2017new}, the smart home~\cite{zeng2017end}, and cryptocurrency systems~\cite{mai2020user}. 
Users create cognitive maps of system components, their relations, and potential privacy risks, which helps them to understand where threats could emerge and how they could take effect.
Research shows that more technically advanced users have a different understanding of digital systems~\cite{kang2015my,gallagher2017new}.
These findings highlight the importance of reasonable technical knowledge for informed user decisions.
In our work, we followed the method of studying mental models in prior literature~\cite{kang2015my} to investigate users' mental models of LLM-based CAs.

\section{Dataset Analysis}

We first analyzed a dataset containing real-world ChatGPT conversations to answer RQ1 regarding real-world ChatGPT users' disclosure behaviors.
In this section, we present the methodology and the findings of the dataset analysis.

\subsection{Methodology}

\subsubsection{The ShareGPT52K Dataset}

We used the ShareGPT52K dataset in our first analysis to examine real-world examples of data sharing practices.
This dataset contains 50,496 ChatGPT chat histories shared by users of the ShareGPT Chrome extension from December 2022 to March 2023.
The conversations from this dataset, each identified by a unique ID, includes both the users’ prompts and ChatGPT’s responses.
The dataset contains conversations that disclose sensitive information (e.g., person's names, personal experiences), as the extension users may not expect the data will be displayed on the website to the public.
Notably, the ShareGPT dataset has become a popular dataset to fine-tune other models~\cite{chiang2023vicuna, zheng2023judging, geng2023koala, gudibande2023false, ji2023towards, mu2023embodiedgpt, zhang2023huatuogpt} in academic research and open-source community, while the risks of handling potentially sensitive personal information shared by users are not well discussed in the literature.

\paragraph{Ethical considerations}
Using this dataset may raise ethical concerns. We believe our use of this dataset is justifiable for two reasons.
1) The primary objective of our research is to comprehend the privacy risks associated with the current disclosure behaviors of ChatGPT users and to identify areas where users need more support to manage these risks.
This research is crucial to prevent incidents like the ShareGPT data leak from recurring.
2) This dataset provides a unique opportunity to closely examine users' sensitive disclosure behaviors. As users were unaware of being observed, they likely exhibited less self-censorship in their shared content. This is akin to the analysis of leaked password datasets in password security literature~\cite{ur2015measuring}.
To minimize the potential harm to individuals included in this dataset, we avoided quoting any text verbatim from the dataset and removed all PII.

\subsubsection{Sampling methods}

We used Microsoft Presidio\footnote{Microsoft Presidio: \url{https://microsoft.github.io/presidio/}} to detect PII in the ShareGPT52K dataset.
This narrowed down our dataset to 30K conversations containing PII.
We further split these 30K conversations into two groups.
One group consisted of 7K conversations that contained, on average, more than one detected PII per turn in the conversation.
The second group consisted of 23K conversations that contained, on average, less than one detected PII per turn in the conversation.
\revision{As our goal was to analyze private-sensitive disclosures, we oversampled responses from the first group. 
To balance the need for a diverse array of cases with the practicality of manual analysis, we randomly selected batches of conversation threads from each group until no new themes emerged and we reached saturation~\cite{guest2006many}.
Hence, for our final dataset, we included a total of 200 conversation threads, each with multiple conversation turns, to be qualitatively coded.
The sample covered conversations of various lengths, measured by turns of conversations ($min=2, max=572, mean=51.9, std=97.0$).}

\subsubsection{Coding process}

The qualitative analysis was guided by the contextual integrity framework~\cite{nissenbaum2004privacy, nissenbaum2020privacy}.
According to the framework, context-relative informational norms are characterized by four key parameters: contexts, actors, data types, and transmission principles.
The transmission principles can be technically analyzed, while the first three parameters depend on the real-world usage of a system.
The primary goal of this analysis is to gain insight into the three parameters that characterize disclosure behaviors in real-world ChatGPT conversations.

First, we analyzed who (\textit{actors}) shared what sensitive information (\textit{data types}) with ChatGPT by re-labeling the detected PII at the conversation level.
To facilitate this task, we created a web-based tool to present a conversation and highlight all the detected occurrences of each PII type.
The coder can use this website to directly label whether each detected PII type appears at least once in the conversation, and label which individuals the detected PIIs were about, with the following options: self (the user who conversed with ChatGPT), others, both (self and others), and unknown.
Two researchers first coded 50 conversations independently and calculated the inter-rater reliability.
They then discussed the discrepancies in their labeling results and created a set of coding criteria (\autoref{sec:pii-coding-criteria}).
We noticed that the automated detection tool resulted in many false positives and a few false negatives.
For false positives, the most common reasons were that the PII data type was misclassified (e.g., a phone number labeled as a passport number), open information about public figures (e.g., Taylor Swift), and made-up examples (e.g., 123-456-7890 as a placeholder for phone number).
The two researchers then labeled another 50 conversations and achieved high inter-rater reliability \revision{(Gwet's AC1)} for frequently detected PII types including \textit{Person Name} \revision{(0.89)}, \textit{Email Address} \revision{(0.87)}, \textit{URL} \revision{(0.88)}, \textit{Date/Time} \revision{(0.81)}, NRP (\textit{Nationality, Religious, or Political Group)} \revision{(0.81)}, \textit{IP Address} \revision{(1)}, \textit{Location} \revision{(0.86)}, \textit{Phone Number} \revision{(0.93)}, \textit{US Bank Number} \revision{(1)}.
Finally, one research coded another 100 conversations.
This coding analysis surfaced 106 conversations out of the 200 conversations that actually contained PII.

Next, we aimed to analyze the \textit{contexts} parameter by creating a typology of sensitive disclosure scenarios.
Two researchers read through the 106 conversations verified to contain PII and wrote extensive summaries of the scenarios. 
The affinity diagramming method was used to organize the different use scenarios based on similar themes\revision{~\cite{beyer1999contextual}}. All conversations were put onto sticky notes. 
After grouping all sticky notes, the researchers iteratively placed labels on the created categories that described the general theme of each group \revision{(\autoref{sec:codebook-dataset-scenarios})}.

\subsubsection{Methodological limitations}

Our method has some limitations.
Firstly, the context of the conversations in the dataset is sometimes unclear, which can lead to uncertainties.
For example, it can be challenging to discern whether a name mentioned by a user is real or fictitious.
Furthermore, users' thought processes when sharing data are not directly observable, which may limit our understanding of their behaviors.
Finally, our dataset has an inherent sample bias: the type of user who was willing to share their conversations with the ShareGPT Chrome extension may not be representative of all LLM users.
Despite these limitations, we have strived to make conservative interpretations of the observed behaviors in the dataset. Additionally, our interview study will provide complementary insights, and the two studies will collectively provide a more comprehensive understanding of the phenomena under investigation.

\subsection{Findings}

\subsubsection{Data Types Disclosed in ChatGPT Conversations}

In our sample, we identified both highly identifiable information such as \textit{Person Name}, \textit{Email Address}, \textit{IP Address}, \textit{Phone Number}, and \textit{Passport Number}, as well as less directly identifiable personal information including \textit{URL}, \textit{Date/Time}, NRP (\textit{Nationality, Religious, or Political Group)}, and \textit{Location}.
We also found many disclosures of sensitive personal experiences that do not directly include typical PII.
Even the more abstract PII types listed above can be used to identify a person given a specific context.
For example, a user appeared to be an elementary school teacher and asked ChatGPT to generate a teaching plan.
The user disclosed information such as the grade they were teaching, as well as the name and the district of the school where they teach.
These two pieces of information, along with the topic area of the teaching plan, might be sufficient to identify the specific person or at least significantly narrow down who it might be.
Furthermore, we found that the PII users shared during these conversations may not always have been necessary for the primary task.
For example, users asked ChatGPT to help fix issues in some code snippets. 
Two phone numbers were found in the code, possibly for testing, which were not needed for the task.

\subsubsection{Actors: Relationship Between the Data Subject and the User}
\label{sec:dataset-others-data-sharing}

Beyond sharing their own PII, we found examples of users disclosing other people's PII in their conversations with ChatGPT.
This finding suggests that in ChatGPT and other LLM-Based CAs, there are not only institutional privacy issues but also interdependent privacy issues.
The disclosure of other individuals' data was common when users used ChatGPT to handle tasks that involved other people  (e.g., friends, colleagues, clients) in both personal and work-related scenarios.
For example, a user shared email conversations regarding complaints about their living conditions and asked ChatGPT about further steps to resolve and go forward with the matter.
The conversation included email communication between the user and a staff member responsible for handling the issue.
It contained both individuals' email addresses, names, and phone numbers.

\subsubsection{Contexts: A Typology of Disclosure Scenarios}
\label{sec:disclosure-scenario-typology}

We developed a multidimensional typology to characterize the disclosure scenarios of ChatGPT.
The typology consists of four dimensions: \textit{Context}, \textit{Topic}, \textit{Purpose}, and \textit{Prompt strategy}.

\textit{Context} includes three categories: Work-Related, Academic-Related, and Life-Related.
Context can affect the data-sharing norms.
For example, a company may not allow its employees to share work-related data with ChatGPT.

\textit{Topic} includes eight categories: Business, Assignment, Programming, Financial, Legal, Medical, Life, and Entertainment (see \autoref{tab:disclosure-typology-topic}).
Some topics are inherently more sensitive, such as the Financial and Medical topics\revision{~\cite{zhao2023survey, pan2020privacy, el2018understanding}}, which naturally lend themselves to users' sharing information like their transaction histories and medical diagnoses, respectively.

\begin{table*}[]
    \centering
    \caption{A Typology of Disclosure Scenarios -- Topic}
    \begin{tabular}{p{0.13\linewidth} p{0.35\linewidth} p{0.45\linewidth}}
    \toprule
    Topic & Examples & Potential Risks\\
    \midrule
    Business & 
\raisebox{-0.8\totalheight}{\includegraphics[width=2.35in]{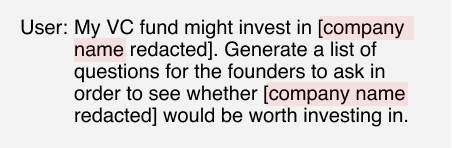}} & Sharing business details like ideas, plans, or strategies can pose privacy risks. This information could be confidential, or potentially reveal sensitive data such as the user's location or company name.
 \\
 Assignment & 
\raisebox{-0.9\totalheight}{\includegraphics[width=2.35in]{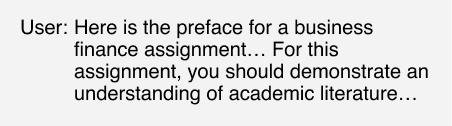}} & Using ChatGPT for assignment could risk plagiarism accusations and inadvertently reveal user information, such as their academic focus.  \\
   Programming      & \raisebox{-0.8\totalheight}{\includegraphics[width=2.35in]{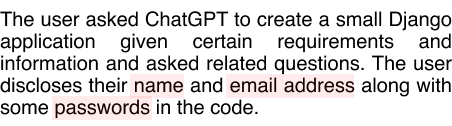}}
 & These scenarios typically pose minimal privacy risks. However, users sometimes incorporate personal details like phone numbers or emails in the shared code.
 \\ 
 Financial & 
\raisebox{-0.85\totalheight}{\includegraphics[width=2.35in]{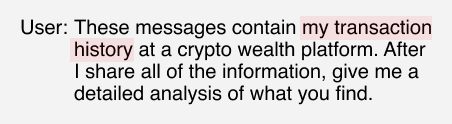}} & Sharing financial details, including transaction histories, might increase the risk of fraudulent activity. \\
Legal & \raisebox{-0.9\totalheight}{\includegraphics[width=2.35in]{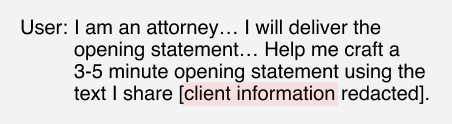}} & Possible privacy risks could involve parties like attorneys, clients, or the legal case. Sharing confidential information with ChatGPT could have serious implications if leaked. \\
Medical & \raisebox{-0.9\totalheight}{\includegraphics[width=2.35in]{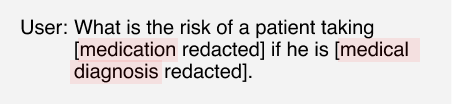}}
 & Possible privacy risks could involve parties like doctors, patients, or the medical case. Sharing confidential information with ChatGPT could have serious implications if leaked.   \\

Life & \raisebox{-0.9\totalheight}{\includegraphics[width=2.35in]{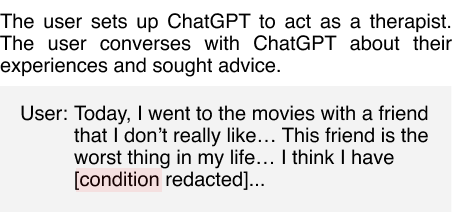}} & Sharing personal information with ChatGPT poses potential privacy risks. Any potential leak could damage reputations or relationships, especially when involving third parties. \\
Entertainment & \raisebox{-0.9\totalheight}{\includegraphics[width=2.35in]{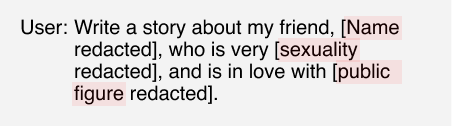}}
 & While much of the content in these scenarios is fictional and thus poses little privacy risk, incorporating personal details can increase this risk. Depending on the scenario type, reputational harm may also occur if users request inappropriate content from ChatGPT.\\
    \bottomrule
    \end{tabular}
    \label{tab:disclosure-typology-topic}
\end{table*}

\textit{Purpose} is the reason users are using ChatGPT, and includes Generating Content, Generating Plans/Advice, Answering Questions, Data Analysis, and Casual Chat.
Conversations in the Data Analysis category were particularly prone to sensitive disclosures: many users directly copied and pasted a data table and asked ChatGPT to derive insights from it.

\textit{Prompt strategy} captures tactical approaches users used to interact with ChatGPT to achieve their goals, including Direct Command, Interactively Defining the Task, Role-Playing, and Jail-Breaking.
Interactively Defining the Task refers to scenarios where users engage in multi-round interactions with ChatGPT.
In these scenarios, users gave ChatGPT instructions and adjusted their instructions based on ChatGPT's response.
This interactive process led users to gradually reveal more information as the conversation progressed; sometimes, this progressive disclosure was driven by ChatGPT.
For example, a user set up ChatGPT to act as a therapist. The user conversed with ChatGPT about their experiences and sought advice through multiple rounds of conversation.
As ChatGPT was giving advice, the user would respond with more specific details, similar to the flow of a conversation between real people.

\subsubsection{Users' Concerns and Protective Behaviors in the Conversations}
\label{sec:user-concerns-dataset}

We found that, in some conversations, users explicitly mentioned having privacy concerns and employed prompt strategies as protective behaviors.
Several users expressed concern --- in their prompts --- that others would find out that they had used AI for the task at hand.
For instance, one user who was writing a book provided ChatGPT with a list of content-generation tasks.
The user explicitly wrote in the prompt like \blockquote{do something for me so that no one will find out this book was written by AI.}.
We also observed users implementing privacy-protective behaviors in their original prompts.
For example, a user asked ChatGPT to help analyze patient messages, and replaced all the names in the messages with ``[PERSONALNAME]''.

 \section{Interview Methodology}
Analyzing the ShareGPT52k dataset allowed us to model in-situ disclosure behaviors (RQ1). Next, to understand when and why users have sensitive disclosure behaviors (RQ2) as well as how users perceive and handle privacy risks (RQ3), we conducted semi-structured interviews with 19 users of LLM-based CAs (18 ChatGPT users, 1 Bing chat user).
The study design was approved by our IRB, and we conducted the interviews remotely between July and August, 2023.
\revision{To refine interview script and recruitment strategy, we conducted pilot studies with users from the authors' social networks, encompassing both technical and non-technical backgrounds. 
After each pilot interview, the researchers reviewed and reflected on the process, took notes, and made adjustments to the interview script.}

\subsection{Participants}

We recruited our participants from Prolific\footnote{\href{https://www.prolific.co}{Prolific} is a website for recruiting research study participants.} and the authors' social network.
To ensure that we had a diverse sample and that we only invited participants with relevant experience to participate in the interview, we used a pre-screening survey that asked about their ChatGPT use experiences, as well as their gender identity, age, and whether they have a technical background.
We learned from pilot interviews that participants with limited experience using LLM-based CAs had little to say in response to our questions, especially with respect to privacy.
Therefore, we mostly selected participants who used ChatGPT or a related LLM-based CA at least once weekly.

We intentionally did not mention ``privacy'' in our recruiting materials to avoid skewing our sample towards more privacy-conscious participants.
Specifically, in the pre-screening survey, we did not directly ask them about what data they have shared with ChatGPT; instead, we designed multiple-choice questions based on the disclosure scenario typology developed from the dataset analysis (\autoref{sec:disclosure-scenario-typology}) to collect information indicative of their data disclosure behaviors.
We ended up recruiting 19 participants with a wide range of use cases, age groups, and technical backgrounds (\autoref{tab:interview-participant-overview}). 
\revision{The interviews were concluded when the research team stopped to learn new insights from new interviews, indicating that data saturation was reached~\cite{guest2006many}, thus ensuring a diverse and comprehensive range of insights.}
All 19 participants completed the main study (around 60 to 90 minutes) and were compensated \$30 USD each.

\begin{table*}[]
    \centering    \caption{Interview Participant Overview}
    \begin{tabular}{p{0.02\linewidth} p{0.05\linewidth} p{0.046\linewidth} p{0.03\linewidth} p{0.05\linewidth} p{0.12\linewidth} p{0.09\linewidth} p{0.42\linewidth}}
    \toprule
    ID & Gender & Age & Tech &  Version & Other services & Frequency & Use cases\\
    \midrule
      P1 & Woman & 18-24 & No &  3.5 & AI Chat & Weekly & Relocation Advice, Career Advice, Schoolwork \\
      P2 & Woman & 18-24 & No &  3.5 & - & Weekly & Career Advice, Schoolwork, Review Writing, Marketing Advice \\
      P3 & Woman & 45-54 & No &  3.5 & - & Weekly & Relocation Advice; Diet Advice, Exercise Advice, Career Advice, Medical Advice, Research Work\\
      P4 & Woman & 25-34 & Yes &  3.5 & - & Not regular & Email Writing, Career Advice, Info search, Email Writing\\
      P5 & Woman & 25-34 & No &  3.5 & - & Weekly & Casual Chat, Math Learning, Email Writing, Copy-editing, Social Media Post Writing\\
      P6 & Woman & 45-54 & Yes &  3.5 & Bard; Pi.ai & Weekly & Finance Advice, Life Advice, Class Preparation, Info search, Test Capabilities\\
      P7 & Woman & 25-34 & No &  3.5 & - & Weekly & Career Advice, Diet Advice, Test Capabilities\\
      P8 & Woman & 25-34 & No &  4; 3.5 & API Playground; Bing chat; Bard; Claude.ai & Daily & Casual Chat, Therapy, Data Analysis, Career Advice, Email Writing, Revise Writing, Programming, Language Learning, Schoolwork, Portfolio Making\\
      P9 & Woman & 25-34 & No &  3.5 & Bard & Weekly & Relocation Advice, Ad Writing, Email Writing, Info Search, Test Capabilities\\
      P10 & Woman & 25-34 & Yes &  4; 3.5 & - & Daily & Casual Chat, Therapy, Data Analysis, Career Advice, Diet Advice, Email Writing, Language Learning\\
      P11 & Man & 25-34 & Yes &  4; 3.5 & API Playground & Daily & Legal Advice, Medical Advice, Programming, Data Analysis, Create Apps, Concepts Learning, Career Advice\\
      P12 & Man & 25-34 & No &  4; 3.5 & Bard & Daily & Finance Advice, Legal Advice, Medical Advice, Book Chapter Writing, Email Writing, Joke Writing, \\
      P13 & Man & 18-24 & Yes &  3.5 & - & Weekly & Casual Chat, Life Advice, Schoolwork, Programming, Data Analysis, Revise Writing, Email/Work Message Writing, Career Advice\\
      P14 & Man & 45-54 & Yes &  4; 3.5 & Bing chat; Pi.ai & Daily & Medical Advice, Finance Advice, \\
      P15 & Man & 65-74 & Yes &  4; 3.5 & - & Daily & Medical Advice, Generate Survey Responses, Social Media Post Writing, \\
      P16 & Man & 45-54 & No &  - & Bing chat & Daily & Therapy,  Casual Chat, Finance Advice\\
      P17 & Man & 25-34 & Yes &  4; 3.5 & - & Daily & Programming, Data Analysis, Immigration Advice, Literature Search, Revise Writing\\
      P18 & Man & 18-24 & Yes &  4; 3.5 & API Playground & Weekly & Programming, Data Analysis, Schoolwork, Revise Writing\\
      P19 & Man & 25-34 & No &  3.5 & - & Not regular & Therapy, Casual Chat, Finance Advice, Career Advice, Diet Advice, Exercise Advice\\
    \bottomrule
    \end{tabular}
    \label{tab:interview-participant-overview}
\end{table*}

\subsection{Study Design}

At the beginning of each interview, the interviewer briefed the participants on study goals, procedures, and data protection measures. The interviewer then obtained the consent for recording (see interview script in \autoref{sec:interview-script}).

Our interview protocol consisted of four parts.
First, we inquired into participants' lived disclosure behaviors and privacy considerations.
Prior to the interview, we asked each participant to prepare at least three conversations with ChatGPT or other LLM-based CAs, redacting any information they did not want us to see.
We encouraged them to select conversations that involved information they considered personal.
During the interview, we first asked them to walk us through these conversations.
For each, we asked them to explain their primary goals, as well as if and why they had any concerns sharing personal data.
For any concerns they mentioned, we asked appropriate follow-up questions, such as if they had taken any measures to address the concerns, if they encountered any challenges, and if they refrained from using LLM-based CAs due to the expressed privacy concerns.

Second, we aimed to understand our participants' mental models of how the LLM-based CAs use their input to generate responses and improve services.
We designed a mental model drawing activity, as has been used in prior work~\cite{kang2015my}.
Participants used the whiteboard feature of Zoom or Google Slides to draw diagrams and also verbally explained their understanding.
We then debriefed them on how the system actually works, using a reference diagram we created.
This diagram was based on the typical LLM inference and training process, as well as specific information from privacy policy of the company that produced the LLM-based CA (e.g., OpenAI, Google).
Then, we asked about users' understanding and feelings about certain topics, including data storage, training, and memorization risks.

Third, we examined users' awareness of and experiences with using existing privacy and data controls in ChatGPT\footnote{For the Bing chat user, we only asked about the history deletion feature in Bing chat.}, including chat history, delete chat, share chat, opt out of having user input used for training models. We also asked about other sensitive practices such as sharing their ChatGPT accounts with other people, and using ChatGPT plugins.

Fourth, we asked whether participants learned anything new or surprising from the interview, whether they wanted to share additional privacy concerns, and whether they had specific requests for improving the existing system design.
Finally, we asked participants to envision and share a wish-list for privacy and data controls for LLM-based CAs.

\subsection{Qualitative Analysis}
We qualitatively analyzed the interview transcripts using a bottom-up open coding method \revision{and concurrently reviewed video recordings of participants reflecting on their chat histories and drawing their mental models.
This approach ensured a more comprehensive understanding of the chatting contexts described by the participant and the mental model diagrams they drew.}
Our analysis involved two rounds of coding as recommended by \citet{saldana2015coding}.

In the first round of coding, two researchers coded the same six interviews independently to develop a codebook.
They held daily meetings to discuss the codes, reconcile coding discrepancies, and iteratively merge their codebooks.
By the end of this round, we derived an initial codebook with 195 codes. Then, the two researchers collectively conducted axial coding to merge similar codes and assign high-level themes for answering the research questions.

In the second round of coding, the remaining 13 interviews were each independently coded by one of the two researchers using the new codebook. Changes were made to the codes and themes as needed, and all changes were discussed and agreed upon by both researchers in daily meetings.
\revision{Following the guidance of~\citet{mcdonald2019reliability}, we did not calculate inter-rater reliability as our goal was to identify emergent themes rather than achieve consensus.}
The final codebook contains 62 codes grouped into 6 themes (see the codebook in \autoref{sec:codebook-interview}). 

\subsection{Methodological Limitations}

Our interview study has some limitations. Firstly, participants may have avoided discussing conversations with particularly sensitive information; moreover, individuals who share more sensitive information with ChatGPT may not have wanted to participate in the study. Accordingly, our sample could be skewed towards less sensitive conversations.
In fact, one person who passed the prescreening requirements decided not to participate in the interview due to their concerns with sharing sensitive data.
Secondly, there may be some ambiguity when participants reported experiences outside of the specific conversations they prepared for the interview, and this could potentially introduce some recall biases.
There is also the possibility of social desirability bias~\cite{grimm2010social}, where participants may tailor their responses to what they perceive the interviewers want to hear.

\section{Interview Results}

In this section, we present our findings from the \interviewsamplesize{} interviews we conducted to answer RQ2 and RQ3.
\revision{Sections 5.2 and 5.3 primarily explore factors influencing users' sensitive disclosure intentions and their trade-off between disclosure benefits and risks, relevant to RQ2.
Sections 5.4, 5.5 and 5.6 examine users' knowledge, awareness, motivations, and the barriers they encounter in adopting threat-protective behaviors, pertaining to RQ3.
However, these aspects often intersect, with some sections relating to multiple research questions.}

\subsection{Overview of Use Cases and Disclosure Behaviors from the Interviews}
\label{sec:findings_overview}

Our participants' conversations covered a wide range of scenarios (\autoref{tab:interview-participant-overview}).
Within these conversations, the types of personal data they shared included PII, personal experiences or conditions (e.g., health, financial, legal conditions), personal thoughts and emotions, and data from other people. 
For example, zip codes were shared for relocation advice (P1\&P9).
Physiological data like age, weight, and height were provided for generating tailored diet and exercise plans (P3\&P8).
Demographic information and medical conditions were shared for medical advice (P3, P11, P12, P14\&P15).
Educational and work experiences were disclosed for career advice or resume revision (P1, P3, P4, P7, P8, P10, P11, P13). 
Writing materials such as emails sent by others (P10) and unpublished papers or books (P2, P12, P13, P17, P18) were shared for content generation, reviewing, or revising purposes.
Some users appeared to develop more emotional relationships with ChatGPT, treating the AI as a ``pen pal'' or a therapist (P8, P10, P13, P15, P16, P19).

\subsection{Factors that Affect Users' Disclosure Intentions with LLM-Based CAs}
\label{sec:factors-affect-disclosure}

We found people generally thought about whether the primary goals of their tasks could be met and whether the operation was convenient enough when deciding what LLM-based CAs to use or what information to share.
However, certain privacy concerns emerged when talking about specific use scenarios.

\subsubsection{Perceived Capability of the CAs (All participants)}
People tend to disclose information if they perceive the CA's capability as beneficial to their task goals. Conversely, they withhold information if they question the agent's competence.
For instance, P15 provided detailed medical diagnosis data after confirming with ChatGPT that it could provide an extensive explanation of the results. 
\label{sec:emotional-support}
Apart from functionality support, LLM-based CAs provided  \textit{emotional support} that encouraged users to disclose information.
Many participants mentioned they treated ChatGPT as a ``friend'' or a ``therapist'', and sometimes had casual chats with it about their lives (P8, P10, P13, P15, P16, P19).
P10 and P19 mentioned that they loved sharing personal life details with ChatGPT due to the positive feedback it gave.
P16 demonstrated a conversation in which he told ChatGPT that he missed his brother, who had passed away, and disclosed a lot of his memories about his brother per ChatGPT's request.
\blockquote[P16]{He asked me to talk to him about my brother. It's like a full conversation. He wanted to know everything.}

\revision{\subsubsection{Convenience of Operation (P1, P2, P4, P8, P10)}
Participants were more inclined to share data when doing so was easily afforded by the interface.
For example, P8 shared PDF documents including original research data with Claude AI\footnote{\url{https://claude.ai/login}} due to the convenience of document input in Claude AI.}
\revision{Conversely, operational barriers deterred some users from sharing.
P10 chose to copy and paste parts of her resume into ChatGPT rather than the entire document due to the inconvenience of inputting long blocks of text. 
\blockquote[P10]{Because there were too many words in my resume. I'm kind of lazy. I don't want to drag from the top to the bottom.}}

\subsubsection{Perceived Personal Data Sensitivity (All participants)}
Participants' perceptions of the specificity and identifiability of personal data influenced their disclosure intention with LLM-based CAs. 
Participants sometimes refrained from sharing data that they considered uniquely identifying.
For example, P3 never shared phone numbers and physical addresses with ChatGPT to avoid being tracked by human reviewers. 
P1 considered her birth date too sensitive to share due to its relation to account security and password resetting.
They were more open to sharing data that could be considered PII, but was not deemed to be uniquely identifying.
For example, P10 was fine with sharing the city in which they resided:
\blockquote[P10]{Telling ChatGPT I live in [city name redacted], it's kind of like, saying I live on the earth.}
Participants also expressed concerns with sharing other types of personal information, even if it was not uniquely identifying.
For instance, P8 preferred to be cautious when sharing personal opinions. 
P3 felt uncomfortable sharing her actual weight.

Notably, the perceived sensitivity of different types of data varies from person to person. For example, P3, P8, P11, and P14 were comfortable sharing their names, while others like P1 and P7 were more guarded. \label{sec:sensitivity_subjective}
Even for the same task, seeking advice for exercise and diet plans, P8 felt ease to share weight data, while P3 chose to provide false information.
In extreme cases, some people expressed concern about sharing any personal information with ChatGPT (P2).

\subsubsection{Resignation (P1, P3, P4, P6, P8, P10, P11, P13, P14)}
People also justified their sharing of personal information by saying their data was already accessible on various platforms such as social media, government databases, and educational systems.
In other words, participants believed that the marginal risk of sharing personal data with LLM-based CAs was low and were resigned to the idea that their individual disclosure decisions had little impact on the accessibility of their personal data.
For instance, P1 equated her data-sharing behavior on ChatGPT with those on social media.
\blockquote[P1]{I'm doing the same risk by using the app like Instagram or Facebook.}

\subsubsection{Perceived Risks and Harms}

\paragraph{Concerns over data misuse by institutions (P2, P3, P5, P6, P8, P9, P11, P12, P13, P14, P17, P18)}
\label{sec:uncertainty-about-data-usage}
Many participants mentioned they were not sure about how their data could be used by OpenAI, and expressed a range of concerns related to potential misuse, encompassing issues such as incomplete data deletion (P6, P9), the possibility of selling user data or using it for marketing (P9, P12), sharing data with third parties (P9, P17), human reviews by OpenAI staff (P3, P6, P14), and public disclosure of data (P6, P9).
On the other hand, some participants (P8, P11, P13, P14) said they trust OpenAI more because they did not think it would sell user data and use that data for marketing purposes. 
P14 recounted experiences with Bing chat, where he faced targeted marketing after specific conversations, an issue he had not encountered with ChatGPT. 
This difference influenced his preference for ChatGPT over Bing chat.

\paragraph{Concerns about others finding out (P8, P17, P18)}
\label{sec:interview-don't-let-others-know-i-used-ai}
Given the uncertainty surrounding the social acceptance of using AI for certain tasks, some people expressed concerns about others finding out that they used ChatGPT. 
For example, P18 did not want his friend to know he used ChatGPT for homework. P8 was worried that others might \blockquote{change their attitude to me} if they discovered her reliance on AI for tasks like schoolwork and email writing.
\blockquote[P8]{I hope they (my professors) will never know I used AI to do that (write emails).}

\paragraph{Concerns about idea theft (P2, P14, P17)}
Users' concerns about idea theft manifested in various ways. These included concerns about the system redistributing their work without acknowledging the author (P2), OpenAI employees seeing and stealing the user's business idea (P14), and allowing other people to read parts of a paper that is under review (P17).
\blockquote[P2]{I don't know if ChatGPT uses it (the fiction that I wrote) as inspiration for other people, or spits it out as it wrote it itself instead of me.}
Note that concerns about sharing original content also vary from person to person. 
For example, P17 was worried about model memorizing and spreading his unpublished work, while P18 willingly shared his unpublished work for paper revision.

\subsubsection{Perceived Norms of Disclosure}

\paragraph{Attitude towards disclosing others' data and having one's own data shared by others (P3, P8, P10, P11, P13, P14, P16)}
\label{sec:interview-others-data-sharing}
Apart from disclosing their own information, participants also discussed their thoughts about disclosing data about others. 
Some people expressed heightened caution when sharing data related to others, even more so than their own data (P3, P11, P13, P14).
P3 and P14 voiced concerns regarding data ownership and the ethics of sharing information without the original owner's consent, noting, \blockquote{That's not my decision to make.}
Conversely, others expressed less concern about sharing information about others (P8, P10, P16). 
P10 also felt minimal concern about her own information being shared by others, viewing it as a fair exchange.
Furthermore, P8 deemed it acceptable to share extensive research data containing details of various individuals, including full names, demographic specifics, personal experiences, and feedback, justifying her actions as being non-commercial in intention.
However, P8 expressed discomfort with the idea of her data being shared by others, worrying about \blockquote{how AI will summarize or do what kind of judgment for me.}.

\paragraph{Caution with sharing work-related data due to company policies or NDAs (P5, P10, P11, P14, P15)}
Several users discussed using ChatGPT for work-related purposes, and expressed caution against sharing company or confidential data, citing company policies or \textit{non-disclosure agreements} (NDAs) as the reason. 
For example, P10 recognized ChatGPT's capability in data analysis but refrained from sharing company datasets due to her company's policies.

\subsection{How Users Navigate the Trade-off Between Disclosure Risks and Benefits}

Users' use of ChatGPT for sensitive tasks surfaces inevitable tensions between privacy, utility, and convenience.
We identified three broad strategies participants used when navigating this trade-off.

\subsubsection{Accept Privacy Risks to Reap Benefits (P4, P8, P9, P10, P14, P15, P18, P19)}

Participants often found accomplishing their objective to be more important than avoiding privacy risks.
As P15 said, \blockquote{There is a price for getting the benefits of using this application}, further adding: \blockquote{It's a fair game}.
Others appeared to be pessimistic about accessing the benefits of LLM-based CAs while also preserving privacy, feeling that \blockquote[P9, P19]{you can't have it both ways}.
For example, P10 used ChatGPT to revise her resume and shared detailed work experiences.
She considered it a necessary trade-off, stating, \blockquote{Let's say I need some advice about resume. If I don't provide those contents that contain a lot of my private things, ChatGPT won't work.}
\label{sec:indispensable-part}
To some participants, ChatGPT has provided irreplaceable value and has become an indispensable part of their life.
For example, P15 said \blockquote{I cannot imagine myself doing my work, my daily activities, without ChatGPT.}

\subsubsection{Avoid Tasks Requiring Personal Data Due to Privacy Concerns (P2, P3, P9, P14, P19)}

Some participants mentioned that their privacy concerns for certain tasks were so significant that they avoided using ChatGPT for those tasks.
For example, P19 once tried to ask ChatGPT for financial advice, and felt it would be really nice if he could provide detailed information like \blockquote{the amount of money that we bring in, the kids that we have.}
However, he also stated, \blockquote{I just never felt comfortable doing that.}
As a result, he only provided generic information, which was less helpful, leading him to terminate the conversation.

\subsubsection{Manually Sanitize Inputs (All but P5, P8, P11, P15)}
\label{sec:ad-hoc-protective-behaviors}

While participants generally employed one of the two strategies above, we found nearly everyone had employed one or more of three ad-hoc privacy-protective measures to try to find a middle ground between privacy and utility when possible.

\paragraph{Censor and/or Falsify Sensitive Information (P1, P3, P6, P7, P9, P10, P12, P13, P16, P19)}
Many participants mentioned that they avoided disclosing sensitive or identifiable information such as their name, social security number, and location.
Participants sometimes chose to only provide coarse-grained or even fake information.
For example, P3 provided a different college name from her alma mater that is in the same university system and asked for career advice, and P13 mentioned he had given ChatGPT fake names and fake information.

\paragraph{Sanitize Inputs Copied from Other Contexts (P4, P17, P18)}
Tasks like copy-editing and programming require users to provide content copied from other contexts.
Some participants mentioned that they post-processed these inputs to sanitize the data.
For example, P4 and P18 removed personal information in emails that they asked ChatGPT to revise and manually added that information back later.
P17 replaced document names included in a PowerShell script with placeholders before sharing it with ChatGPT.
In addition to redacting sensitive information, P17 also mentioned limiting the amount of information he shared with ChatGPT in any one prompt.
He used ChatGPT to proofread his paper and only copied one or a small number of sentences each time due to concerns that ChatGPT might remember the entire paper.

\paragraph{Only Seek General Advice (P2, P10, P14, P17)}
Some participants only sought general advice, trading the specificity of the response they received for improved privacy.
For example, P14 mentioned he only felt comfortable having general conversations with ChatGPT about business.
This strategy can require users to spend more time summarizing the task, making their conversations not only less specific but also less convenient.
For example, P10 mentioned that she used ChatGPT to help with data analysis tasks at work.
Since she was not permitted to share the raw data, she needed to summarize the data schema to share with ChatGPT, rather than directly asking ChatGPT to generate the code based on the data.

\subsection{Mental Models of How LLM-Based CAs Handle User Input}
We summarize users' mental models for two processes: response generation (Model A: ``ChatGPT is magic'', B: ``ChatGPT is a super searcher'', C: ``ChatGPT is a stochastic parrot'') and model improvement and training (Model D: ``User input is a quality indicator'', E: ``User input is training data'').
We found that users' mental models demonstrated overly simplified or flawed understandings of how their data was used and how LLM-based CAs worked.

\subsubsection{Mental Models of Response Generation}

\paragraph{Model A: ChatGPT is magic (P8, P10, P12, P15)}
\begin{figure}
    \centering
    \includegraphics[width=0.8\linewidth]{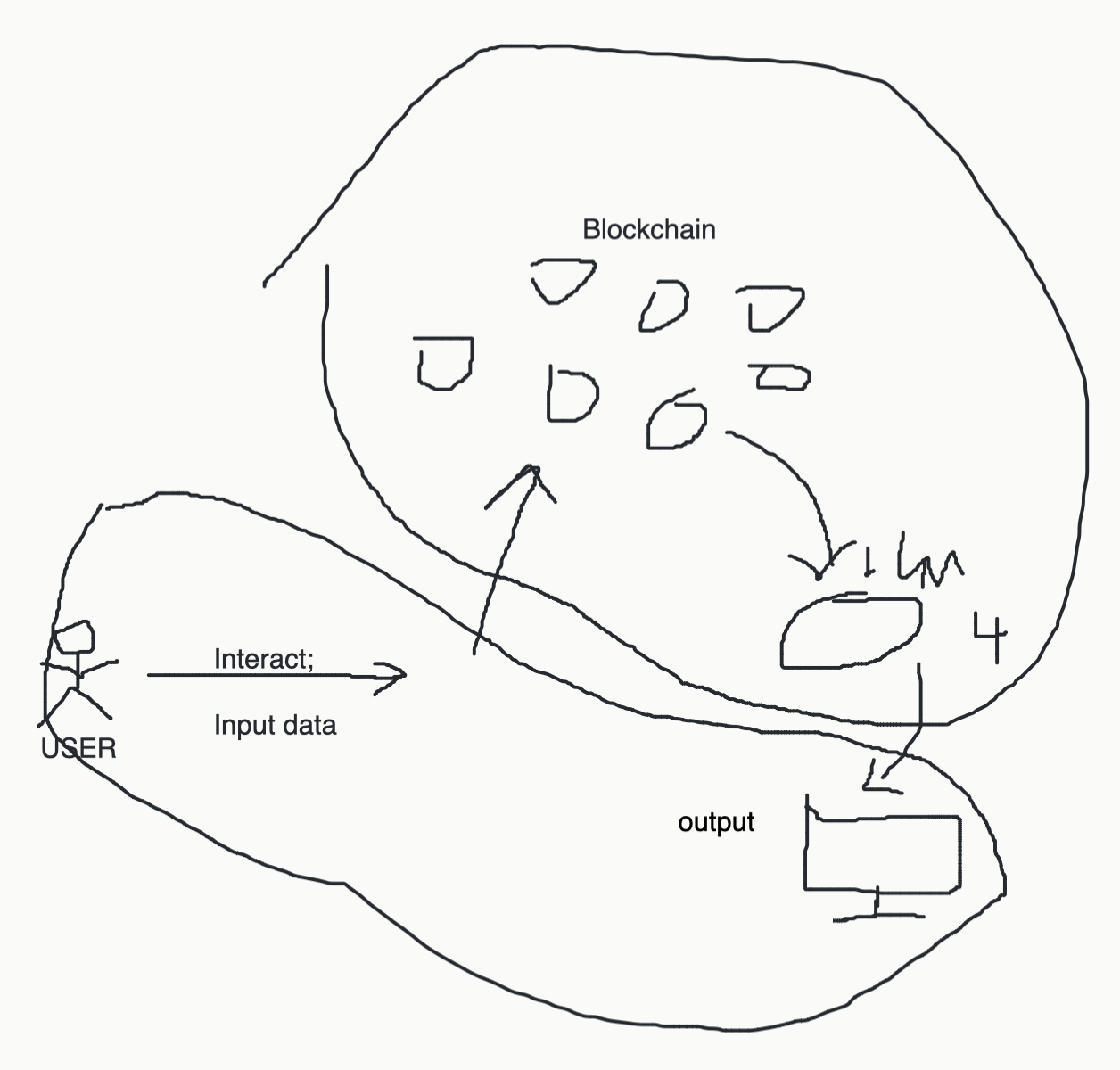}
    \Description[Mental model drawing]{This figure illustrates a drawing created by a participant with mental model A who thought the ChatGPT is an AI Blackbox. On the left of the image is a small figure, symbolizing the user. Arrows indicate data flow: starting from the small figure (user) with an arrow pointing right labeled "interact" and "input data", leading to multiple small circles labeled "blockchain". From there, another arrow points to a larger oval labeled "LLM 4", which then points to a computer labeled "output". A large circle encompasses the user figure and computer, denoting local processes, while another encircles the blockchain circles and oval, indicating remote processes.}
    \caption{Screenshot of P8's drawing representing mental model A: ChatGPT is magic.}
    \label{fig:P8MentalModel}
\end{figure}

Mental model A represents a shallow technical understanding of how ChatGPT generates responses. 
Participants who harbored this mental model thought of the generation process as an abstract transaction: messages are sent to an LLM or a database, and an output is received.
P8 illustrated a typical example of this model, shown in ~\autoref{fig:P8MentalModel}. 
In her words:
\blockquote{ChatGPT uses the computing power to generate something to send to the LLM, the model of ChatGPT. And then you get your output data...Actually it likes a blackbox for me. I just use it. I mean, I never thought about that before.}
Similarly, P10 described the response generation process as \blockquote{some kinds of magic I don't know}.

\paragraph{Model B: ChatGPT is a super searcher (P1, P2, P3, P4, P5, P7, P16, P19)}
\begin{figure*}
    \centering
    \includegraphics[width=1\linewidth]{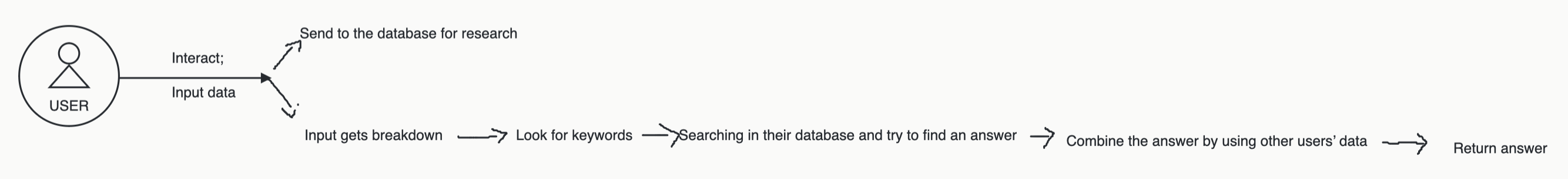}
    \Description[Mental model drawing]{This figure illustrates a drawing created by a participant with mental model B who thought the ChatGPT searches for relevant information and synthesizes the results into text response. On the left, a small figure symbolizes the user. Data flow is shown by arrows: from the user, one arrow extends right with labels "interact" and "input data". It bifurcates into two: the top arrow pointing right-top reads "send to the database for research" and the bottom one progresses rightward with labels showing the sequence: "input gets broken down" -> "look for keywords" -> "search database for answers" -> "combine answer using others' data" -> "return answer".
}
    \caption{Screenshot of P4's drawing representing mental model B: ChatGPT is a super searcher.}
    \label{fig:P4MentalModel}
\end{figure*}

Participants with this mental model often envisioned the response generation process as a form of live keyword search on the internet or in a database sourced from the internet, followed by a synthesis of the gathered information. 
As shown in \autoref{fig:P4MentalModel}, P4 generally described the generation process as 
\blockquote{My input will get broken down. And then look for keywords...After the keywords, will start searching in their database, trying to find an answer. At this point they might try to combine the answer.}
Participants with this mental model often expected rule-based methods or human interventions to play a role in generating the responses. 
For example, P19 envisioned rules that match keywords to databases pulled from the internet. 
P16 assumed that there were humans-in-the-loop in generating responses. 
Apart from the internet-at-large, a few participants noted that the system might include other data resources. 
For example, P7 believed users' conversations would be included in ChatGPT's knowledge base.

\paragraph{Model C. ChatGPT is a stochastic parrot (P6, P11, P13, P14, P17, P18)}
\begin{figure}
    \centering
    \includegraphics[width=0.8\linewidth]{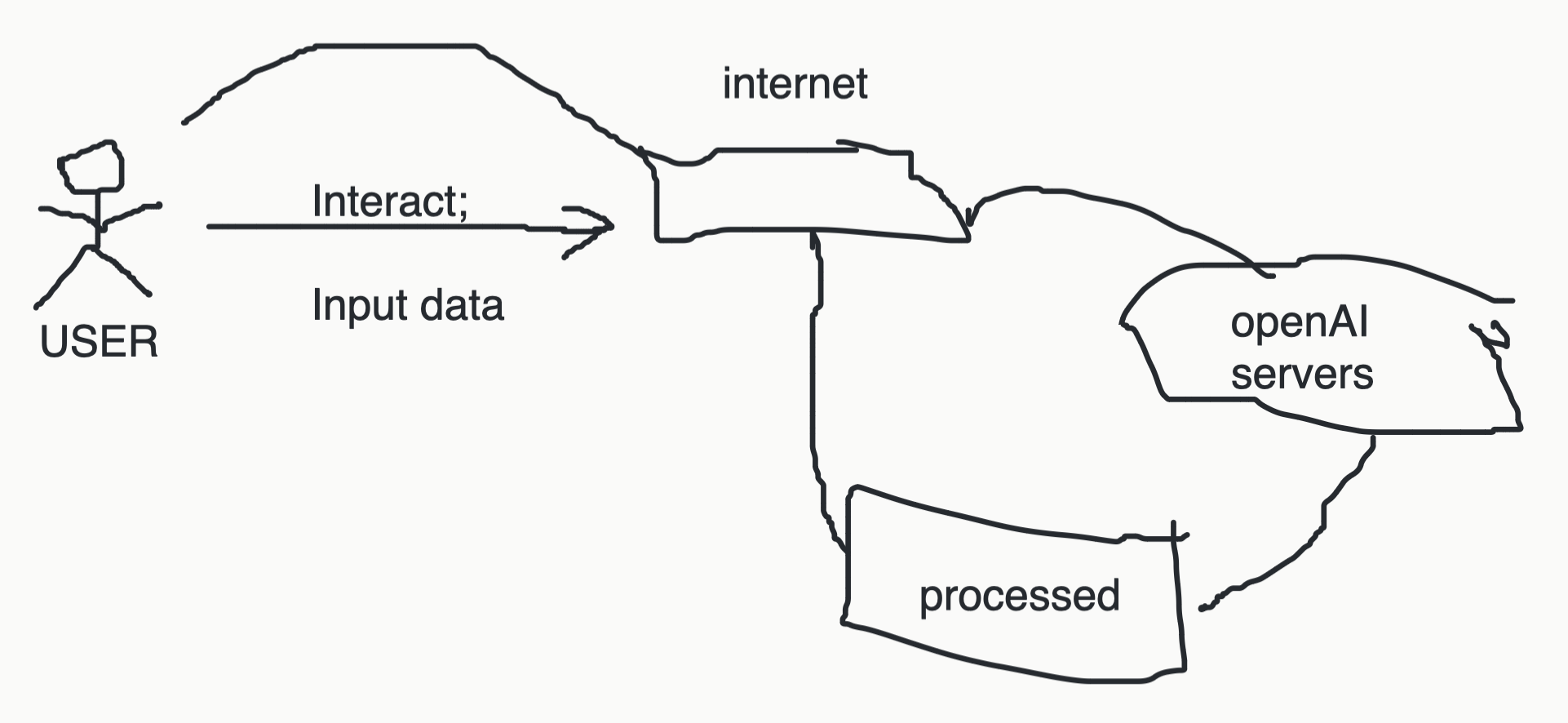}
    \Description[Mental model drawing]{This figure illustrates a drawing created by a participant with mental model C who thought the ChatGPT uses an end-to-end machine learning model for response generation. On the left, a small figure represents the user. Arrows show data flow: starting with the user, an arrow labeled "interact" and "input data" points to a rectangle symbolizing the internet. This internet rectangle connects to two other rectangles: "processed" and "OpenAI servers," forming a loop among the three. Finally, an arrow from the internet rectangle loops back to the user, denoting the returned responses.}
    \caption{Screenshot of P14's drawing representing mental model C: ChatGPT is a stochastic parrot. P14 verbally explained how the end-to-end machine learning model generates a response in technical detail.}
    \label{fig:P14MentaModel}
\end{figure}

Participants with Model C articulated a more sophisticated understanding of the response generation process.
They believed that the response generation was handled by an end-to-end machine learning model that stochastically generated each word, in sequence, based on user input and previously generated words.
Unlike Model B, users with Model C did not expect the system to have separate components for understanding the query, gathering related information, and generating the reply; instead, they understood that the response was generated by a single, trained model. 
For example, P14 drew \autoref{fig:P14MentaModel} and verbalized, \blockquote{It (my input data) would go through the Internet, then goes to their servers...They’re just getting a big block of vectors, and makes its predictive actions, does all of its what you know processing, and then it produces a response.}
All participants with this mental model had technical backgrounds.

\subsubsection{Mental Models on Improvement and Training}

\paragraph{Model D: User input is a quality indicator (P1, P3, P4, P6, P9, P10, P15, P19)}
Users with Mental Model D believed that their inputs were used to assess their satisfaction with responses, and that the system would learn over time to produce responses more similar to those that were rated highly. 
Others felt their inputs were used as a semantic key to help index similar questions. 
Thus, if someone asks a similar question, the LLM can select and respond with answers to similar questions that were rated more positively.
\label{sec:hard-to-imagine-memorization}
Because they felt that user input was isolated from system output, participants with this model were less able to expect and understand memorization risks.
For example, P1 could not see how the personal data mentioned in her prompt, such as zip code, can be used in generating future responses. 
P3 believed that human reviewers might label previous responses as good or bad, and determine if those responses can be reused in the future.

This mental model resembles the reinforcement learning from human feedback (RLHF) method that OpenAI used to train the InstructGPT model~\cite{ouyang2022training}.
However, with RLHF, the context presented in the prompt may also be memorized and regenerated by future models, contradicting the expectations of people who hold this mental model.

\paragraph{Model E: User input is training data (P2, P5, P7, P11, P12, P13, P14, P16, P17, P18).}
Users with this type of mental model understood their their input prompts could be used to influence future responses.
For example, P11 said if his input data had been used for training, \blockquote{there's a possibility that's going to use the data as an output for somebody else.}
Some people with this type of mental model had a technical background (P11, P13, P14, P17, P18).
Others without a technical background did not know the specific training process that could be used, but they expected that their information could be reused in future responses to other users.

We found participants had different expectations as to whether their inputs are used to improve a global model accessible to all users (P7, P11, P12, P13, P14, P16, P17, P18) or a personalized model used only by themselves (P2, P5).
P5 believed that the model is personalized to each user based on their inputs. 
This was because she experienced significantly improved responses from ChatGPT within a conversation thread. 
Similarly, P2 was hesitant to share her ChatGPT account with others because she believed the model was personalized and that inputs from others could adversely affect its training.

\subsection{Users' Awareness and Reactions to Memorization Risks in LLM-based CAs}
\label{sec:findings_memorization_risks}

Prior work has shown that LLMs can memorize training data and leak this training data as a response to the right prompt~\cite{carlini2021extracting, carlini2022quantifying}. This memorization effect entails unique privacy risks for using LLM-based CAs that utilize user data to improve their models.
However, most participants lacked awareness of this issue, and only P2, P17 and P18 brought up this issue before we explicitly asked about it.
P18 thought memorization risks allow for new privacy attacks, and P2 and P17 brought it up as a concern that had limited their data sharing with ChatGPT for real-world use cases.
After we explained memorization risks to the other 16 participants, most participants remained unconcerned. 
Four participants (P5, P12, P13, P19) expressed surprise and heightened concern about disclosing personal data to ChatGPT: especially data about emotional states, work-related information, and sensitive PII like social security numbers.
Three of them stated that they might alter their data-sharing behaviors in the future.

\subsubsection{Concerned with Memorization: Prior Exposure to Leaks (P17)}
P17 personally encountered a memorization leak when using Copilot, which made him concerned about the possibility of his own personal information being memorized by ChatGPT. 
\blockquote[P17]{When the Copilot plugin for VS Code was published, I installed it. I typed the name of my classmate and it auto completed my classmate's school ID. It's very terrible.}

\subsubsection{Concerned with Memorization: Intellectual Property Leaks  (P2, P17)}
\label{sec:concerns-aboutmemorization-risks}
P2 and P17 expressed concerns ChatGPT memorizing and distributing their original writing without credit, notice, or consent. 
P2 hesitated to use ChatGPT to review her short stories, worrying that ChatGPT might generate new content based on her original work or inspire other users with her work.
Similarly, P17 was cautious when sharing unpublished paper content and limited the input in each session to prevent the system from disclosing the whole paper or key ideas to other users before publication.

\subsubsection{Unconcerned with Memorization: Does Not Share Sensitive Data (P1, P6, P7, P8, P9, P10, P11, P16)}
After being introduced to memorization risks when using ChatGPT, most participants expressed minimal concerns, primarily because they had not shared data they deemed sensitive, or because they believed even the sensitive data being memorized was not linked to their identities. 
For instance, P9 was not particularly worried because she has been cautious and has not shared personal information beyond her age and the city where she lives. 
P6 was unconcerned because she believed any sensitive data memorized by the AI would not be linked to her identity.
\blockquote[P6]{It may remember some things, but I assume that it would disassociate my specific identity with what it memorized.}

\subsubsection{Unconcerned with Memorization: The Risk is Too Abstract (P1, P8, P10)}
\label{sec:unconcerned-because-not-undersatnding-memorization}
Some participants expressed limited concern about memorization risks because the risks were difficult to comprehend. 
The difficulty was tied to their mental model of how ChatGPT improves AI performance based on their input. 
For example, both P1 and P10 held Mental Model D\footnote{We were not able to obtain a clear answer from P8 about her mental model about improvement and training.}, and believed that ChatGPT treated their input data strictly as a quality indicator rather than as training data. This mental model, in turn, made it hard to imagine memorization leaks (see \autoref{sec:hard-to-imagine-memorization}). 
The absence of prior exposure to memorization leaks also added to their difficulty in imagining memorization leaks. \blockquote[P8]{I don't have this kind of experiences like my information was shown to others as output. And when I imagine it, I'm not so confident about it...I haven't seen this kind of news before.}

\subsubsection{Unconcerned with Memorization: Plausible Deniability of AI-Generated Output (P3, P8)}
Some participants felt unconcerned because the model would not produce accurate information about them (P3), and believed that others may not perceive the output as accurate information (P8). 
For example, P3 was indifferent about the potential disclosure of her weight --- which she considered to be sensitive data --- since she had provided ChatGPT with incorrect information. 
P8 was not concerned due to a similar reason.
\blockquote[P8]{Although I use ChatGPT for many tasks, I'm not concerned about my input being released to other people, because they can't judge if it's fake information.}

\subsection{Users' Awareness, Understanding, and Attitudes About Existing Privacy Controls}

Finally, we examined how users utilize existing support to protect their privacy and how they would like the system to be improved to make them feel safer.

\subsubsection{Users Lack Awareness of Existing Privacy Controls (All but P11, P14, P16, P17, P18)}
ChatGPT provides a control that allows users to opt out of having their conversations used by OpenAI for training, negating memorization risks.
However, most participants had not heard about this control.
After learning about it, most participants felt it was a good feature.
\label{sec:opt-out-reactions}
Many participants expressed a desire to use it even though they did not feel concerned about what they shared with ChatGPT.
For example, P2 said \blockquote{even though I know there's nothing bad that's gonna happen. But still, I would just want the peace of mind of being able to opt out.}

\subsubsection{Dark Patterns Impeded Adoption of the Opt-out Control (All participants)}
\label{sec:dark-patterns}

\begin{figure*}
    \centering
     \begin{subfigure}[b]{0.48\textwidth}
         \centering
          \includegraphics[width=\textwidth]{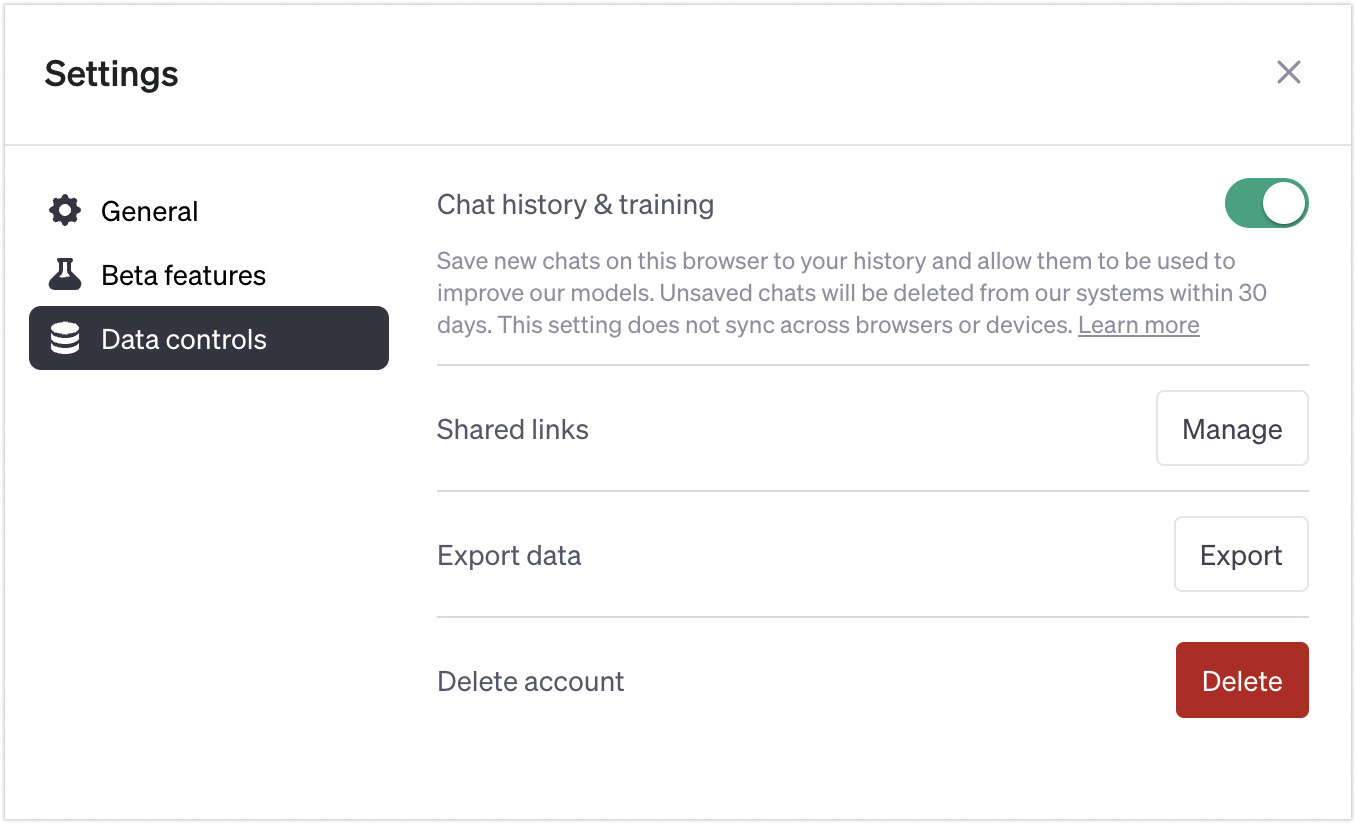}
         \caption{Chat history \& training bundled together (easier to discover)}
         \label{fig:opt-out-history-settings}
     \end{subfigure}
     \hfill
     \begin{subfigure}[b]{0.48\textwidth}
         \centering
         \includegraphics[width=\textwidth]{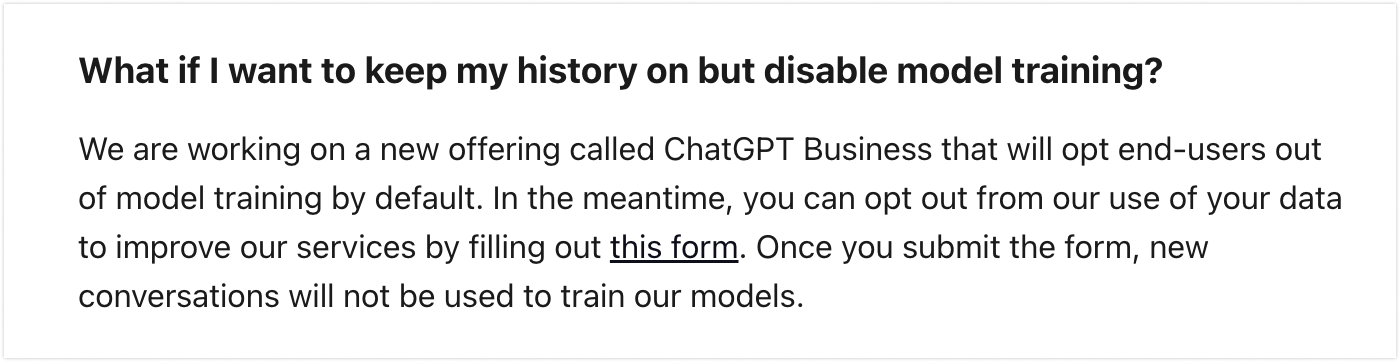}
         \caption{Turning off training and keeping history (harder to discover)}
         \label{fig:opt-out-form}
     \end{subfigure}
     \hfill
     \Description[Two screenshots of dark patterns in ChatGPT]{The two figures illustrate two methods for opting out, with image (a) highlighting an easily accessible method via ChatGPT’s interface and image (b) a more obscure method found in OpenAI’s Data Control FAQ. (a) On the left, the ChatGPT settings pop-up is displayed with "Data controls" selected in the navigation. At the top is the "Chat history \& training" header accompanied by an activated switch for Opt-in. The subsequent text details that saved chats are retained for history and may be used to enhance models, while chats not saved are deleted after 30 days. A clickable "Learn more" link is also present. (b) On the right, a section of OpenAI’s Data Control FAQ is displayed. Titled "What if I want to keep my history on but disable model training?" it mentions the upcoming ChatGPT Business offering which defaults to opting users out of model training. Meanwhile, users can opt out by completing a linked form, ensuring that new conversations won’t be used for training.}
    \caption{Dark patterns in ChatGPT: ChatGPT offers two ways for a user to opt out of having their data used for model training. The one in the user settings is easier to discover (all but P15 found it), while the training and chat history opt-out control are bundled together, so a user who wants to opt out of model training will have to turn off the chat history feature as well. The users could also submit a form to opt out of training and keep the history, while it is in an \href{https://help.openai.com/en/articles/7730893-data-controls-faq}{FAQ article} that is harder to discover (none of our participants found it). \revision{The above issues were observed during our studies in August 2023. As of November 2023, the form link is still in the FAQ article, while this form has been disabled and it further directs users to the OpenAI Privacy Request Portal to submit privacy requests. As of February 2023, the form link is replaced with the link to the privacy portal.}}
    \label{fig:opt-out-dark-pattern}
\end{figure*}

We identified multiple dark patterns in ChatGPT's opt-out design that hindered users' adoption of this feature.
First, ChatGPT uses user data for training by default.
This surprised many participants, especially for P17 who was a paid ChatGPT Plus user.
\blockquote[P17]{I think OpenAI should not use my data (for model training), because I paid for my ChatGPT.}

Second, there are two ways for users to opt out of having their data used for model training (\autoref{fig:opt-out-dark-pattern}).
One was easier to discover (all but P15 found it), but it forces users to turn off storing chat history at the same time (\autoref{fig:opt-out-history-settings}).
Several participants said they were hesitant (P11) or did not want (P8, P7, P17, P18) to opt out of training because they wanted to keep the history feature.
\blockquote[P8]{I'm a little annoyed about that...Because I still want to keep my history, I need to provide my information for them to train. It's like something mandatory.}
Alternatively, users can also submit a form to opt out of training and keep the history feature (\autoref{fig:opt-out-form}), but the link to access this form is hidden in a FAQ article\footnote{Data Controls FAQ: \url{https://help.openai.com/en/articles/7730893-data-controls-faq}}.
None of our participants found this option independently.
The inconvenience seemed to further discourage users from enabling the opt-out feature, as put by P2, \blockquote{To be honest, if it's not easy to find, I feel like the form will be complicated. So yeah, I would probably just not (fill it out).}

\subsubsection{Users Anticipated More Granular Opt-out Controls (P1, P6, P8, P13, P14, P18)}
\label{sec:granular-opt-out}

Several participants shared that the opt-out control worked at the conversation-level, and not the account level whereby some conversations could be specially designated as ``opt out'': i.e., they expected behavior similar to a browser's incognito mode.
This understanding was shared by both people with (P6, P13, P14, P18) and without (P1, P8) a technical background.
P14 hoped to have this more granular control not for privacy but for improving the quality of training data.
He often tested the capability of ChatGPT with \blockquote{all kinds of wild things}, and he said \blockquote{I would think that the appropriate thing to do is that anything communicated during an opt-out is not added to the training data...because that would just make the AI go insane.}

\section{Discussion}
In this section, we present the user-driven privacy threats synthesized from our results, and then establish guidelines for LLM privacy research in both the short and long term.

\subsection{User-Driven Privacy Threat Modeling of LLM-Based Conversational Agents} 

Our results grounded the technical privacy risks of LLM such as the memorization risks~\cite{carlini2021extracting, carlini2022quantifying} in the contexts of users' actual interactions with ChatGPT.
We also provided empirical evidence supporting potential risks caused by LLMs, as speculated in prior literature~\cite{weidinger2021ethical}.
In the following, we summarize privacy threats around this new technology that users are concerned about and may cause harm to users under specific use cases.

\subsubsection{User Concerns About Institutional Privacy}

Much as expected, institutional privacy issues were repeatedly brought up by our participants. 
Most participants are concerned about traditional privacy risks that do not solely exist in LLM-based CAs, such as incomplete data deletion and having user data reviewed by humans (\autoref{sec:uncertainty-about-data-usage}).
Fewer participants raised LLM-specific concerns such as memorization risks (\autoref{sec:concerns-aboutmemorization-risks}).
\revision{It is essential to note that users' trust in ChatGPT has largely been influenced by the fact that it currently does not profit from user profiling, advertising, and selling data.}
However, since this technology is still in its nascent phase, we are likely to see it adopted in many more fields, accommodated by various business models.
The use of rich conversational data for targeted advertising or marketing isn't unimaginable, given the lack of clear regulations and the tempting potential, and might cause bigger concerns.

\subsubsection{User Concerns About Interdependent Privacy}

We also observed that users use LLM-based CAs to handle tasks that involve other people’s data (\autoref{sec:dataset-others-data-sharing}) and they have concerns about it (\autoref{sec:interview-others-data-sharing}).
Specifically, the novel use cases enabled by LLM-based CAs (e.g., asking ChatGPT to draft a response to colleague's emails) allow people to share unprecedented types and amounts of information about \textit{other people} compared to similar systems (e.g., search engine, traditional CAs).
Our results show that there are severe interdependent privacy issues in LLM-based CAs, which are even harder to address since most current models for privacy management are heavily individualized.

\subsubsection{The Model is One-Size-Fits-All, but User Concerns Are Not}

A recurring theme of the interview results is that users' privacy concerns are contextualized and subjective (\autoref{sec:factors-affect-disclosure}).
Users' reasons about why or why not sharing certain data with ChatGPT involved all the key parameters of the contextual integrity framework~\cite{nissenbaum2004privacy, nissenbaum2020privacy}.
Despite varied use cases and expectations, only one end-to-end model is serving all the requests in these LLM-based systems.
This means that the system works in a way that is agnostic to the sensitivity of use cases and the varied norms about data collection, sharing, and retention under each use case.
As a result, the burden of navigating the varying privacy risks essentially falls back to users.
Our results suggest that people take various ad-hoc approaches to desensitize their input (\autoref{sec:ad-hoc-protective-behaviors}).
However, these methods can be tedious (e.g., redacting all the PII in an email), and users may not always remember to apply them (e.g., one participant forgot he was not permitted to share his work-related data with ChatGPT).

\subsubsection{The Impact of Human-Like Interactions on Privacy}

Work in robot-human interaction has shown that people relate to non-human agents in social ways, including extending moral judgments to them~\cite{kahn2012people}, attributing them sociocultural awareness~\cite{simmons2011believable}, and trusting them based on their behavior and anthropomorphism~\cite{natarajan2020effects} as well as, most relevantly for our work, their use of language~\cite{ye2023improved}.
While past work has emphasized the benefits of this trust for collaboration, our results suggest that it can also lead to the gradually increasing disclosure of sensitive information. Such human-like interactions may also act as a nudge that affects what types of information the user shares (\autoref{sec:disclosure-scenario-typology} and \autoref{sec:emotional-support}).
Future work needs to distinguish between increased trust as a benefit for collaboration and the dangers of unwarranted trust from insufficient transparency and the leveraging of human social cognition leading to greater privacy harms.

\subsubsection{The Emerging Fear of Being Found Out Using AI}
The lack of established norms regarding when it is appropriate or not to use AI has raised concerns among users of ChatGPT.
Some use cases, such as using AI to generate book chapters, are indeed inappropriate (\autoref{sec:user-concerns-dataset}).
Others are more benign but may lead to people questioning the user's ability because they used AI to assist with their work. (e.g., a non-native English speaker using AI for polishing writing, \autoref{sec:interview-don't-let-others-know-i-used-ai}).
It remains to be discussed how to balance individual privacy protection and the societal impact in these situations.

\subsection{Design Privacy-Friendly LLM-Based Conversational Agents and Other LLM-Based Applications}

Our studies suggest that the developers of LLM-based conversational agents or other applications should make more efforts to ensure the system is designed in the best interest of the users, and provide sufficient support for users to navigate the risks and benefits under different contexts (see our typology of disclosure behaviors \autoref{sec:disclosure-scenario-typology}).
We propose potential improvement directions below. 

\subsubsection{Consider User Mental Models for Designing Privacy Support}

The types of mental models identified in our studies suggest that systems based on LLM often do not function as users anticipate. This suggests that system design should consider this mismatch, adopting an intuitive design that matches users' expectations or proactively communicating the privacy risks that users may not foresee.
Users' mental models might also be affected by the frontend and interaction design.
For example, P4 felt GitHub Copilot was safer than ChatGPT because \blockquote{that's offline}.
However, Copilot is supported by the OpenAI Codex model, which is also hosted on the cloud.
It suggests that when LLMs are more deeply integrated into a system that feels private (e.g., the IDE), it may be difficult for users to understand that their data will be sent off the device.
Furthermore, it might become harder for users to understand which part of their data remains on the device and which part gets sent out if the system employs a hybrid design combining local and remote models.

\subsubsection{Assist Users in Taking Privacy-Protective Measures}

Many users are taking certain privacy-protective measures, such as omitting or obfuscating sensitive information, telling privacy lies, and segmenting input, while it could become tiring to manually desensitize all the input, and users may forget to do it.
Therefore, there is an opportunity for designing privacy-preserving techniques that assist users in applying these measures in an automated or semi-automated manner.
For example, a smaller model dedicated to detecting PII and other more complex disclosures may be distilled from a large model fine-tuned for the task, and can be run locally to either remind users of the sensitive information included in their input or automatically rewrite it before sending it out.
Overall, there needs to be more research in designing and evaluating techniques that help users achieve utility while maintaining their privacy.

\subsubsection{Be Cautious About Using User Data for Training Models}
The lack of understanding of the model training process and awareness of memorization risks suggest that LLM-based systems should exercise caution when using user data for model training.
If such data is used, it is crucial to communicate the associated risks to users effectively.
It is also important to provide users with convenient and granular control, enabling them to easily opt-in or out according to their best interests (\autoref{sec:granular-opt-out}).
We want to highlight that many participants had positive attitudes towards contributing their data for improving the system,
and that increased transparency could also lead to improved data quality.

\subsubsection{Using Local LLMs for Building Apps in Specific Domains}

Lastly, we want to note that most participants with a technical background believed that only an offline model could completely eliminate their concerns.
Although there is still a significant gap in the performance between smaller, open-source models that can run on devices and the gigantic, proprietary models that run on the cloud, there may be certain use cases where a smaller model may suffice.
The developers of LLM-based systems should keep in mind the option of using a local model whenever it is possible.

\subsection{``It's a Fair Game'', or Is It? Considering the Impact of Erroneous Mental Models, Transparency, Trust, and the Nature of Current LLMs on Prospects for Privacy in LLMs}

Our results reveal challenging problems that may not have a clear direction for resolution, given the privacy models in current technology, law, and social literature.

\subsubsection{Current LLMs are Inherently Surveillant}
Today's LLMs are only as capable as they are due to the sheer scale of data used in training them. This, we argue, marries them structurally to the modern internet's surveillance-as-default mode of operation, with its imperative to always collect more data to train models with the goal of monetizing behavioral surplus~\cite{zuboff2023age}.
Therefore, we caution against the expectation that guidelines encouraging privacy-preserving design (such as those we provide above) will be meaningfully adopted by the makers of LLMs, given the demonstrated tendency of companies operating under these imperatives to reduce privacy to mere compliance and public relations~\cite{waldman2021industry}.
After all, if everyone were to opt out of data collection, and if LLMs were to be meaningfully cautious in how they used public data for training, models of modern levels of sophistication could not exist.
If models are to continue to be trained, some paradigmatic shifts will be needed to understand how such scales of data can be collected and used ethically.

\subsubsection{Transparency and Mental Models}
In laying out a human-centered research roadmap for transparency in LLMs, \citet{liao2023ai} define transparency as ``enabling relevant stakeholders to form an appropriate understanding of a model or system’s capabilities, limitations, how it works, and how to use or control its outputs''. Our results showed that participants exhibited serious, privacy-relevant misconceptions about core concepts including how LLMs generate responses and how user data is used for training models, suggesting the goal is still far away from being achieved. 
Our findings around participants' mental models of LLMs suggest that transparency isn't in good shape for end users.
The task of educating the mass about how LLM-based systems function remains a significant challenge.
However, it is a prerequisite for the ethical deployment of such systems in society.

\subsubsection{Trust and Control}

Extensive research on privacy law and the privacy paradox has demonstrated that not only cognitive biases (i.e., trust and dark patterns) but also the sheer scale, lack of transparency, and binary nature of privacy choices online leads people to quite rationally give up on attempting to engage in privacy self-management~\cite{solove2021myth,hargittai2016can}.
The uniquely social way in which people relate to CAs, our results suggest, may only make an individually choice-driven approach to privacy even more intractable.
Hence, we argue for caution in believing that incremental improvements to privacy controls will have broad effectiveness for LLM privacy.

The other side of this coin, of course, is that neither participants' weak enforcement of privacy boundaries nor their initial assertions about a lack of privacy concerns should be interpreted to suggest that they do not value privacy generally or with LLM-based CAs in particular.
Our interview studies have revealed patterns in users' behaviors that may be explained as a ``Reverse Privacy Paradox''~\cite{colnago2023there}, in which individuals who seem to disregard privacy concerns still engage in privacy-seeking behaviors (\autoref{sec:ad-hoc-protective-behaviors} and \autoref{sec:opt-out-reactions}).
However, these privacy-protective behaviors are not well supported by the system, or even obstructed by it due to the dark patterns (\autoref{sec:dark-patterns}), likely due to the imperative to collect data for training models discussed above.

\subsection{Limitations and Future Work}

In this work, we take the first step towards modeling end-user disclosure behaviors, mental models, and the privacy concerns thereof with LLM-based conversational agents.
There are still some important questions that cannot be fully answered in this work, and we would like to leave them for future research.
\revision{First, the study primarily focused on ChatGPT due to its broad user base and data availability. This focus may limit our understanding of privacy concerns on other LLM-based platforms with different interaction styles or user demographics.}
Second, the interview targeted the general population. However, we speculate that the opportunity cost of not using ChatGPT might be higher for certain populations, such as older adults, non-native speakers, and therapy seekers.
And more research is needed to understand the privacy issues for specific vulnerable populations. 
Third, in this study, our qualitative findings suggest multiple factors that may affect users' risk perceptions and privacy-seeking behaviors related to LLM-based CAs, such as contexts, folk models, and user awareness of privacy controls.
However, more research is needed to quantitatively model the effect of these factors to guide future system design.
Lastly, as LLM is a new technology and users' mental models are still evolving, it is important to conduct longitudinal studies that measure how public attitudes towards privacy issues in LLM-based systems change over time.

\section{Conclusion}
\revision{In this work, we studied and distilled LLM privacy concerns from the perspective of end-users. This knowledge is crucial to advance ongoing debates around AI regulation, as well for HCI researchers seeking to design privacy controls that help end-users address these concerns.}
We first qualitatively analyzed the ShareGPT52K dataset, uncovering various sensitive disclosure behaviors of ChatGPT users in their use of the CA.
Then we conducted semi-structured interviews with 19 users of LLM-based CAs (e.g., ChatGPT) to inquire users about their disclosure behaviors contextualized in their real-world chat logs with LLM-based CAs.
Our results suggest that ChatGPT users want to protect their privacy when possible, while contending with the perceived trade-offs of both convenience and utility.
We found that the one-size-fits-all nature of the model underlying LLM-based CAs largely places the responsibility of protecting privacy on the users.
This is challenging for users due to flawed mental models and and the dark patterns in ChatGPT's opt-out interface.
\revision{Our user-centered investigation has revealed a host of novel problems that require attention from the HCI and LLM research communities.}
We also highlight complex issues and structural problems that require paradigmatic shifts in technology, law, and society.

\begin{acks}
This project is in part supported by a CMU CyLab Seed Funding.
\end{acks}

\bibliographystyle{ACM-Reference-Format}
\bibliography{sample-base}


\begin{thebibliography}{81}


\ifx \showCODEN    \undefined \def \showCODEN     #1{\unskip}     \fi
\ifx \showDOI      \undefined \def \showDOI       #1{#1}\fi
\ifx \showISBNx    \undefined \def \showISBNx     #1{\unskip}     \fi
\ifx \showISBNxiii \undefined \def \showISBNxiii  #1{\unskip}     \fi
\ifx \showISSN     \undefined \def \showISSN      #1{\unskip}     \fi
\ifx \showLCCN     \undefined \def \showLCCN      #1{\unskip}     \fi
\ifx \shownote     \undefined \def \shownote      #1{#1}          \fi
\ifx \showarticletitle \undefined \def \showarticletitle #1{#1}   \fi
\ifx \showURL      \undefined \def \showURL       {\relax}        \fi
\providecommand\bibfield[2]{#2}
\providecommand\bibinfo[2]{#2}
\providecommand\natexlab[1]{#1}
\providecommand\showeprint[2][]{arXiv:#2}

\bibitem[Anderson et~al\mbox{.}(2020)]%
        {anderson2020mental}
\bibfield{author}{\bibinfo{person}{Andrew Anderson}, \bibinfo{person}{Jonathan
  Dodge}, \bibinfo{person}{Amrita Sadarangani}, \bibinfo{person}{Zoe
  Juozapaitis}, \bibinfo{person}{Evan Newman}, \bibinfo{person}{Jed Irvine},
  \bibinfo{person}{Souti Chattopadhyay}, \bibinfo{person}{Matthew Olson},
  \bibinfo{person}{Alan Fern}, {and} \bibinfo{person}{Margaret Burnett}.}
  \bibinfo{year}{2020}\natexlab{}.
\newblock \showarticletitle{Mental models of mere mortals with explanations of
  reinforcement learning}.
\newblock \bibinfo{journal}{\emph{ACM Transactions on Interactive Intelligent
  Systems (TiiS)}} \bibinfo{volume}{10}, \bibinfo{number}{2}
  (\bibinfo{year}{2020}), \bibinfo{pages}{1--37}.
\newblock


\bibitem[Andrews et~al\mbox{.}(2023)]%
        {andrews2023role}
\bibfield{author}{\bibinfo{person}{Robert~W Andrews}, \bibinfo{person}{J~Mason
  Lilly}, \bibinfo{person}{Divya Srivastava}, {and} \bibinfo{person}{Karen~M
  Feigh}.} \bibinfo{year}{2023}\natexlab{}.
\newblock \showarticletitle{The role of shared mental models in human-AI teams:
  a theoretical review}.
\newblock \bibinfo{journal}{\emph{Theoretical Issues in Ergonomics Science}}
  \bibinfo{volume}{24}, \bibinfo{number}{2} (\bibinfo{year}{2023}),
  \bibinfo{pages}{129--175}.
\newblock


\bibitem[Anell et~al\mbox{.}(2020)]%
        {anell2020end}
\bibfield{author}{\bibinfo{person}{Simon Anell}, \bibinfo{person}{Lea
  Gr{\"o}ber}, {and} \bibinfo{person}{Katharina Krombholz}.}
  \bibinfo{year}{2020}\natexlab{}.
\newblock \showarticletitle{End user and expert perceptions of threats and
  potential countermeasures}. In \bibinfo{booktitle}{\emph{2020 IEEE European
  Symposium on Security and Privacy Workshops (EuroS\&PW)}}. IEEE,
  \bibinfo{pages}{230--239}.
\newblock


\bibitem[Bansal et~al\mbox{.}(2019)]%
        {bansal2019beyond}
\bibfield{author}{\bibinfo{person}{Gagan Bansal}, \bibinfo{person}{Besmira
  Nushi}, \bibinfo{person}{Ece Kamar}, \bibinfo{person}{Walter~S Lasecki},
  \bibinfo{person}{Daniel~S Weld}, {and} \bibinfo{person}{Eric Horvitz}.}
  \bibinfo{year}{2019}\natexlab{}.
\newblock \showarticletitle{Beyond accuracy: The role of mental models in
  human-AI team performance}. In \bibinfo{booktitle}{\emph{Proceedings of the
  AAAI conference on human computation and crowdsourcing}},
  Vol.~\bibinfo{volume}{7}. \bibinfo{pages}{2--11}.
\newblock


\bibitem[Beyer and Holtzblatt(1999)]%
        {beyer1999contextual}
\bibfield{author}{\bibinfo{person}{Hugh Beyer} {and} \bibinfo{person}{Karen
  Holtzblatt}.} \bibinfo{year}{1999}\natexlab{}.
\newblock \showarticletitle{Contextual design}.
\newblock \bibinfo{journal}{\emph{interactions}} \bibinfo{volume}{6},
  \bibinfo{number}{1} (\bibinfo{year}{1999}), \bibinfo{pages}{32--42}.
\newblock


\bibitem[Bieringer et~al\mbox{.}(2022)]%
        {bieringer2022industrial}
\bibfield{author}{\bibinfo{person}{Lukas Bieringer}, \bibinfo{person}{Kathrin
  Grosse}, \bibinfo{person}{Michael Backes}, \bibinfo{person}{Battista Biggio},
  {and} \bibinfo{person}{Katharina Krombholz}.}
  \bibinfo{year}{2022}\natexlab{}.
\newblock \showarticletitle{Industrial practitioners' mental models of
  adversarial machine learning}. In \bibinfo{booktitle}{\emph{Eighteenth
  Symposium on Usable Privacy and Security (SOUPS 2022)}}.
  \bibinfo{pages}{97--116}.
\newblock


\bibitem[Brown et~al\mbox{.}(2022)]%
        {brown2022does}
\bibfield{author}{\bibinfo{person}{Hannah Brown}, \bibinfo{person}{Katherine
  Lee}, \bibinfo{person}{Fatemehsadat Mireshghallah}, \bibinfo{person}{Reza
  Shokri}, {and} \bibinfo{person}{Florian Tram{\`e}r}.}
  \bibinfo{year}{2022}\natexlab{}.
\newblock \showarticletitle{What does it mean for a language model to preserve
  privacy?}. In \bibinfo{booktitle}{\emph{Proceedings of the 2022 ACM
  Conference on Fairness, Accountability, and Transparency}}.
  \bibinfo{pages}{2280--2292}.
\newblock


\bibitem[Carlini et~al\mbox{.}(2022)]%
        {carlini2022quantifying}
\bibfield{author}{\bibinfo{person}{Nicholas Carlini}, \bibinfo{person}{Daphne
  Ippolito}, \bibinfo{person}{Matthew Jagielski}, \bibinfo{person}{Katherine
  Lee}, \bibinfo{person}{Florian Tramer}, {and} \bibinfo{person}{Chiyuan
  Zhang}.} \bibinfo{year}{2022}\natexlab{}.
\newblock \showarticletitle{Quantifying memorization across neural language
  models}.
\newblock \bibinfo{journal}{\emph{arXiv preprint arXiv:2202.07646}}
  (\bibinfo{year}{2022}).
\newblock


\bibitem[Carlini et~al\mbox{.}(2021)]%
        {carlini2021extracting}
\bibfield{author}{\bibinfo{person}{Nicholas Carlini}, \bibinfo{person}{Florian
  Tramer}, \bibinfo{person}{Eric Wallace}, \bibinfo{person}{Matthew Jagielski},
  \bibinfo{person}{Ariel Herbert-Voss}, \bibinfo{person}{Katherine Lee},
  \bibinfo{person}{Adam Roberts}, \bibinfo{person}{Tom~B Brown},
  \bibinfo{person}{Dawn Song}, \bibinfo{person}{Ulfar Erlingsson},
  {et~al\mbox{.}}} \bibinfo{year}{2021}\natexlab{}.
\newblock \showarticletitle{Extracting Training Data from Large Language
  Models.}. In \bibinfo{booktitle}{\emph{USENIX Security Symposium}},
  Vol.~\bibinfo{volume}{6}.
\newblock


\bibitem[Chiang et~al\mbox{.}(2023)]%
        {chiang2023vicuna}
\bibfield{author}{\bibinfo{person}{Wei-Lin Chiang}, \bibinfo{person}{Zhuohan
  Li}, \bibinfo{person}{Zi Lin}, \bibinfo{person}{Ying Sheng},
  \bibinfo{person}{Zhanghao Wu}, \bibinfo{person}{Hao Zhang},
  \bibinfo{person}{Lianmin Zheng}, \bibinfo{person}{Siyuan Zhuang},
  \bibinfo{person}{Yonghao Zhuang}, \bibinfo{person}{Joseph~E Gonzalez},
  {et~al\mbox{.}}} \bibinfo{year}{2023}\natexlab{}.
\newblock \showarticletitle{Vicuna: An open-source chatbot impressing gpt-4
  with 90\%* chatgpt quality}.
\newblock \bibinfo{journal}{\emph{See https://vicuna. lmsys. org (accessed 14
  April 2023)}} (\bibinfo{year}{2023}).
\newblock


\bibitem[Colnago et~al\mbox{.}(2023)]%
        {colnago2023there}
\bibfield{author}{\bibinfo{person}{Jessica Colnago},
  \bibinfo{person}{Lorrie~Faith Cranor}, {and} \bibinfo{person}{Alessandro
  Acquisti}.} \bibinfo{year}{2023}\natexlab{}.
\newblock \showarticletitle{Is there a reverse privacy paradox? an exploratory
  analysis of gaps between privacy perspectives and privacy-seeking behaviors}.
\newblock \bibinfo{journal}{\emph{Proceedings on Privacy Enhancing
  Technologies}}  \bibinfo{volume}{1} (\bibinfo{year}{2023}),
  \bibinfo{pages}{455--476}.
\newblock


\bibitem[Cropanzano and Mitchell(2005)]%
        {cropanzano2005social}
\bibfield{author}{\bibinfo{person}{Russell Cropanzano} {and}
  \bibinfo{person}{Marie~S Mitchell}.} \bibinfo{year}{2005}\natexlab{}.
\newblock \showarticletitle{Social exchange theory: An interdisciplinary
  review}.
\newblock \bibinfo{journal}{\emph{Journal of management}} \bibinfo{volume}{31},
  \bibinfo{number}{6} (\bibinfo{year}{2005}), \bibinfo{pages}{874--900}.
\newblock


\bibitem[Dupuy et~al\mbox{.}(2022)]%
        {dupuy2022efficient}
\bibfield{author}{\bibinfo{person}{Christophe Dupuy}, \bibinfo{person}{Radhika
  Arava}, \bibinfo{person}{Rahul Gupta}, {and} \bibinfo{person}{Anna
  Rumshisky}.} \bibinfo{year}{2022}\natexlab{}.
\newblock \showarticletitle{An efficient dp-sgd mechanism for large scale nlu
  models}. In \bibinfo{booktitle}{\emph{ICASSP 2022-2022 IEEE International
  Conference on Acoustics, Speech and Signal Processing (ICASSP)}}. IEEE,
  \bibinfo{pages}{4118--4122}.
\newblock


\bibitem[Dwyer et~al\mbox{.}(2007)]%
        {dwyer2007trust}
\bibfield{author}{\bibinfo{person}{Catherine Dwyer}, \bibinfo{person}{Starr
  Hiltz}, {and} \bibinfo{person}{Katia Passerini}.}
  \bibinfo{year}{2007}\natexlab{}.
\newblock \showarticletitle{Trust and privacy concern within social networking
  sites: A comparison of Facebook and MySpace}.
\newblock \bibinfo{journal}{\emph{AMCIS 2007 proceedings}}
  (\bibinfo{year}{2007}), \bibinfo{pages}{339}.
\newblock


\bibitem[El~Haddad et~al\mbox{.}(2018)]%
        {el2018understanding}
\bibfield{author}{\bibinfo{person}{Ghada El~Haddad}, \bibinfo{person}{Esma
  Aimeur}, {and} \bibinfo{person}{Hicham Hage}.}
  \bibinfo{year}{2018}\natexlab{}.
\newblock \showarticletitle{Understanding trust, privacy and financial fears in
  online payment}. In \bibinfo{booktitle}{\emph{2018 17th IEEE International
  Conference On Trust, Security And Privacy In Computing And
  Communications/12th IEEE International Conference On Big Data Science And
  Engineering (TrustCom/BigDataSE)}}. IEEE, \bibinfo{pages}{28--36}.
\newblock


\bibitem[Estrada(2023)]%
        {Estrada2023}
\bibfield{author}{\bibinfo{person}{Sheryl Estrada}.}
  \bibinfo{year}{2023}\natexlab{}.
\newblock \bibinfo{title}{A startup CFO used ChatGPT to build an FP\&A
  tool—here’s how it went}.
\newblock
\newblock
\urldef\tempurl%
\url{https://fortune.com/2023/03/01/startup-cfo-chatgpt-finance-tool/}
\showURL{%
\tempurl}
\newblock
\shownote{Accessed: 09/11/2023}.


\bibitem[Ferreira(2023)]%
        {Ferreira2023}
\bibfield{author}{\bibinfo{person}{Pedro Ferreira}.}
  \bibinfo{year}{2023}\natexlab{}.
\newblock \bibinfo{title}{Can ChatGPT Improve Technical Analysis and Trading
  Techniques?}
\newblock
\newblock
\urldef\tempurl%
\url{https://www.financemagnates.com/trending/can-chatgpt-improve-technical-analysis-and-trading-techniques/}
\showURL{%
\tempurl}
\newblock
\shownote{Accessed: 09/11/2023}.


\bibitem[Fox(2023)]%
        {Fox2023}
\bibfield{author}{\bibinfo{person}{Andrea Fox}.}
  \bibinfo{year}{2023}\natexlab{}.
\newblock \bibinfo{title}{ChatGPT scored 72
  study shows}.
\newblock
\newblock
\urldef\tempurl%
\url{https://www.healthcareitnews.com/news/chatgpt-scored-72-clinical-decision-accuracy-mgb-study-shows}
\showURL{%
\tempurl}
\newblock
\shownote{Accessed: 09/11/2023}.


\bibitem[Gallagher et~al\mbox{.}(2017)]%
        {gallagher2017new}
\bibfield{author}{\bibinfo{person}{Kevin Gallagher}, \bibinfo{person}{Sameer
  Patil}, {and} \bibinfo{person}{Nasir Memon}.}
  \bibinfo{year}{2017}\natexlab{}.
\newblock \showarticletitle{New Me: Understanding Expert and $\{$Non-Expert$\}$
  Perceptions and Usage of the Tor Anonymity Network}. In
  \bibinfo{booktitle}{\emph{Thirteenth Symposium on Usable Privacy and Security
  (SOUPS 2017)}}. \bibinfo{pages}{385--398}.
\newblock


\bibitem[Geng et~al\mbox{.}(2023)]%
        {geng2023koala}
\bibfield{author}{\bibinfo{person}{Xinyang Geng}, \bibinfo{person}{Arnav
  Gudibande}, \bibinfo{person}{Hao Liu}, \bibinfo{person}{Eric Wallace},
  \bibinfo{person}{Pieter Abbeel}, \bibinfo{person}{Sergey Levine}, {and}
  \bibinfo{person}{Dawn Song}.} \bibinfo{year}{2023}\natexlab{}.
\newblock \showarticletitle{Koala: A dialogue model for academic research}.
\newblock \bibinfo{journal}{\emph{Blog post, April}}  \bibinfo{volume}{1}
  (\bibinfo{year}{2023}).
\newblock


\bibitem[Germain(2023)]%
        {Germain2023}
\bibfield{author}{\bibinfo{person}{Thomas Germain}.}
  \bibinfo{year}{2023}\natexlab{}.
\newblock \bibinfo{title}{A Mental Health App Tested ChatGPT on Its Users. The
  Founder Said Backlash Was Just a Misunderstanding.}
\newblock
\newblock
\urldef\tempurl%
\url{https://gizmodo.com/mental-health-therapy-app-ai-koko-chatgpt-rob-morris-1849965534/}
\showURL{%
\tempurl}
\newblock
\shownote{Accessed: 09/11/2023}.


\bibitem[Gibbs et~al\mbox{.}(2011)]%
        {gibbs2011first}
\bibfield{author}{\bibinfo{person}{Jennifer~L Gibbs}, \bibinfo{person}{Nicole~B
  Ellison}, {and} \bibinfo{person}{Chih-Hui Lai}.}
  \bibinfo{year}{2011}\natexlab{}.
\newblock \showarticletitle{First comes love, then comes Google: An
  investigation of uncertainty reduction strategies and self-disclosure in
  online dating}.
\newblock \bibinfo{journal}{\emph{Communication Research}}
  \bibinfo{volume}{38}, \bibinfo{number}{1} (\bibinfo{year}{2011}),
  \bibinfo{pages}{70--100}.
\newblock


\bibitem[Grimm(2010)]%
        {grimm2010social}
\bibfield{author}{\bibinfo{person}{Pamela Grimm}.}
  \bibinfo{year}{2010}\natexlab{}.
\newblock \showarticletitle{Social desirability bias}.
\newblock \bibinfo{journal}{\emph{Wiley international encyclopedia of
  marketing}} (\bibinfo{year}{2010}).
\newblock


\bibitem[Grodzinsky and Tavani(2010)]%
        {grodzinsky2010applying}
\bibfield{author}{\bibinfo{person}{Frances Grodzinsky} {and}
  \bibinfo{person}{Herman~T Tavani}.} \bibinfo{year}{2010}\natexlab{}.
\newblock \showarticletitle{Applying the “contextual integrity” model of
  privacy to personal blogs in the blogoshere}.
\newblock  (\bibinfo{year}{2010}).
\newblock


\bibitem[Gudibande et~al\mbox{.}(2023)]%
        {gudibande2023false}
\bibfield{author}{\bibinfo{person}{Arnav Gudibande}, \bibinfo{person}{Eric
  Wallace}, \bibinfo{person}{Charlie Snell}, \bibinfo{person}{Xinyang Geng},
  \bibinfo{person}{Hao Liu}, \bibinfo{person}{Pieter Abbeel},
  \bibinfo{person}{Sergey Levine}, {and} \bibinfo{person}{Dawn Song}.}
  \bibinfo{year}{2023}\natexlab{}.
\newblock \showarticletitle{The False Promise of Imitating Proprietary LLMs}.
\newblock \bibinfo{journal}{\emph{arXiv preprint arXiv:2305.15717}}
  (\bibinfo{year}{2023}).
\newblock


\bibitem[Guest et~al\mbox{.}(2006)]%
        {guest2006many}
\bibfield{author}{\bibinfo{person}{Greg Guest}, \bibinfo{person}{Arwen Bunce},
  {and} \bibinfo{person}{Laura Johnson}.} \bibinfo{year}{2006}\natexlab{}.
\newblock \showarticletitle{How many interviews are enough? An experiment with
  data saturation and variability}.
\newblock \bibinfo{journal}{\emph{Field methods}} \bibinfo{volume}{18},
  \bibinfo{number}{1} (\bibinfo{year}{2006}), \bibinfo{pages}{59--82}.
\newblock


\bibitem[Hargittai and Marwick(2016)]%
        {hargittai2016can}
\bibfield{author}{\bibinfo{person}{Eszter Hargittai} {and}
  \bibinfo{person}{Alice Marwick}.} \bibinfo{year}{2016}\natexlab{}.
\newblock \showarticletitle{“What can I really do?” Explaining the privacy
  paradox with online apathy}.
\newblock \bibinfo{journal}{\emph{International journal of communication}}
  \bibinfo{volume}{10} (\bibinfo{year}{2016}), \bibinfo{pages}{21}.
\newblock


\bibitem[Hitron et~al\mbox{.}(2019)]%
        {hitron2019can}
\bibfield{author}{\bibinfo{person}{Tom Hitron}, \bibinfo{person}{Yoav Orlev},
  \bibinfo{person}{Iddo Wald}, \bibinfo{person}{Ariel Shamir},
  \bibinfo{person}{Hadas Erel}, {and} \bibinfo{person}{Oren Zuckerman}.}
  \bibinfo{year}{2019}\natexlab{}.
\newblock \showarticletitle{Can children understand machine learning concepts?
  The effect of uncovering black boxes}. In
  \bibinfo{booktitle}{\emph{Proceedings of the 2019 CHI conference on human
  factors in computing systems}}. \bibinfo{pages}{1--11}.
\newblock


\bibitem[Ischen et~al\mbox{.}(2020)]%
        {ischen2020privacy}
\bibfield{author}{\bibinfo{person}{Carolin Ischen}, \bibinfo{person}{Theo
  Araujo}, \bibinfo{person}{Hilde Voorveld}, \bibinfo{person}{Guda van Noort},
  {and} \bibinfo{person}{Edith Smit}.} \bibinfo{year}{2020}\natexlab{}.
\newblock \bibinfo{booktitle}{\emph{Privacy Concerns in Chatbot Interactions}}.
\newblock \bibinfo{publisher}{Springer International Publishing},
  \bibinfo{pages}{34â€“48}.
\newblock
\showISBNx{9783030395407}
\showISSN{1611-3349}
\urldef\tempurl%
\url{https://doi.org/10.1007/978-3-030-39540-7_3}
\showDOI{\tempurl}


\bibitem[Jang et~al\mbox{.}(2022)]%
        {jang2022knowledge}
\bibfield{author}{\bibinfo{person}{Joel Jang}, \bibinfo{person}{Dongkeun Yoon},
  \bibinfo{person}{Sohee Yang}, \bibinfo{person}{Sungmin Cha},
  \bibinfo{person}{Moontae Lee}, \bibinfo{person}{Lajanugen Logeswaran}, {and}
  \bibinfo{person}{Minjoon Seo}.} \bibinfo{year}{2022}\natexlab{}.
\newblock \showarticletitle{Knowledge unlearning for mitigating privacy risks
  in language models}.
\newblock \bibinfo{journal}{\emph{arXiv preprint arXiv:2210.01504}}
  (\bibinfo{year}{2022}).
\newblock


\bibitem[Ji et~al\mbox{.}(2023)]%
        {ji2023towards}
\bibfield{author}{\bibinfo{person}{Yunjie Ji}, \bibinfo{person}{Yan Gong},
  \bibinfo{person}{Yong Deng}, \bibinfo{person}{Yiping Peng},
  \bibinfo{person}{Qiang Niu}, \bibinfo{person}{Baochang Ma}, {and}
  \bibinfo{person}{Xiangang Li}.} \bibinfo{year}{2023}\natexlab{}.
\newblock \showarticletitle{Towards Better Instruction Following Language
  Models for Chinese: Investigating the Impact of Training Data and
  Evaluation}.
\newblock \bibinfo{journal}{\emph{arXiv preprint arXiv:2304.07854}}
  (\bibinfo{year}{2023}).
\newblock


\bibitem[Kahn~Jr et~al\mbox{.}(2012)]%
        {kahn2012people}
\bibfield{author}{\bibinfo{person}{Peter~H Kahn~Jr}, \bibinfo{person}{Takayuki
  Kanda}, \bibinfo{person}{Hiroshi Ishiguro}, \bibinfo{person}{Brian~T Gill},
  \bibinfo{person}{Jolina~H Ruckert}, \bibinfo{person}{Solace Shen},
  \bibinfo{person}{Heather~E Gary}, \bibinfo{person}{Aimee~L Reichert},
  \bibinfo{person}{Nathan~G Freier}, {and} \bibinfo{person}{Rachel~L
  Severson}.} \bibinfo{year}{2012}\natexlab{}.
\newblock \showarticletitle{Do people hold a humanoid robot morally accountable
  for the harm it causes?}. In \bibinfo{booktitle}{\emph{Proceedings of the
  seventh annual ACM/IEEE international conference on Human-Robot
  Interaction}}. \bibinfo{pages}{33--40}.
\newblock


\bibitem[Kandpal et~al\mbox{.}(2022)]%
        {kandpal2022deduplicating}
\bibfield{author}{\bibinfo{person}{Nikhil Kandpal}, \bibinfo{person}{Eric
  Wallace}, {and} \bibinfo{person}{Colin Raffel}.}
  \bibinfo{year}{2022}\natexlab{}.
\newblock \showarticletitle{Deduplicating training data mitigates privacy risks
  in language models}. In \bibinfo{booktitle}{\emph{International Conference on
  Machine Learning}}. PMLR, \bibinfo{pages}{10697--10707}.
\newblock


\bibitem[Kang et~al\mbox{.}(2015)]%
        {kang2015my}
\bibfield{author}{\bibinfo{person}{Ruogu Kang}, \bibinfo{person}{Laura
  Dabbish}, \bibinfo{person}{Nathaniel Fruchter}, {and} \bibinfo{person}{Sara
  Kiesler}.} \bibinfo{year}{2015}\natexlab{}.
\newblock \showarticletitle{$\{$“My$\}$ Data Just Goes $\{$Everywhere:”$\}$
  User Mental Models of the Internet and Implications for Privacy and
  Security}. In \bibinfo{booktitle}{\emph{Eleventh symposium on usable privacy
  and security (SOUPS 2015)}}. \bibinfo{pages}{39--52}.
\newblock


\bibitem[Kaur et~al\mbox{.}(2020)]%
        {kaur2020interpreting}
\bibfield{author}{\bibinfo{person}{Harmanpreet Kaur}, \bibinfo{person}{Harsha
  Nori}, \bibinfo{person}{Samuel Jenkins}, \bibinfo{person}{Rich Caruana},
  \bibinfo{person}{Hanna Wallach}, {and} \bibinfo{person}{Jennifer
  Wortman~Vaughan}.} \bibinfo{year}{2020}\natexlab{}.
\newblock \showarticletitle{Interpreting interpretability: understanding data
  scientists' use of interpretability tools for machine learning}. In
  \bibinfo{booktitle}{\emph{Proceedings of the 2020 CHI conference on human
  factors in computing systems}}. \bibinfo{pages}{1--14}.
\newblock


\bibitem[Kim et~al\mbox{.}(2023)]%
        {kim2023propile}
\bibfield{author}{\bibinfo{person}{Siwon Kim}, \bibinfo{person}{Sangdoo Yun},
  \bibinfo{person}{Hwaran Lee}, \bibinfo{person}{Martin Gubri},
  \bibinfo{person}{Sungroh Yoon}, {and} \bibinfo{person}{Seong~Joon Oh}.}
  \bibinfo{year}{2023}\natexlab{}.
\newblock \showarticletitle{ProPILE: Probing Privacy Leakage in Large Language
  Models}.
\newblock \bibinfo{journal}{\emph{arXiv preprint arXiv:2307.01881}}
  (\bibinfo{year}{2023}).
\newblock


\bibitem[Kim and Sundar(2012)]%
        {kim2012anthropomorphism}
\bibfield{author}{\bibinfo{person}{Youjeong Kim} {and} \bibinfo{person}{S~Shyam
  Sundar}.} \bibinfo{year}{2012}\natexlab{}.
\newblock \showarticletitle{Anthropomorphism of computers: Is it mindful or
  mindless?}
\newblock \bibinfo{journal}{\emph{Computers in Human Behavior}}
  \bibinfo{volume}{28}, \bibinfo{number}{1} (\bibinfo{year}{2012}),
  \bibinfo{pages}{241--250}.
\newblock


\bibitem[Kimmel(2023)]%
        {kimmel2023chatgpt}
\bibfield{author}{\bibinfo{person}{Daniel Kimmel}.}
  \bibinfo{year}{2023}\natexlab{}.
\newblock \bibinfo{title}{ChatGPT Therapy Is Good, But It Misses What Makes Us
  Human}.
\newblock
  \bibinfo{howpublished}{\url{https://www.columbiapsychiatry.org/news/chatgpt-therapy-is-good-but-it-misses-what-makes-us-human}}.
\newblock
\newblock
\shownote{Accessed: 09/11/2023}.


\bibitem[Kshetri(2023)]%
        {kshetri2023cybercrime}
\bibfield{author}{\bibinfo{person}{Nir Kshetri}.}
  \bibinfo{year}{2023}\natexlab{}.
\newblock \showarticletitle{Cybercrime and privacy threats of large language
  models}.
\newblock \bibinfo{journal}{\emph{IT Professional}} \bibinfo{volume}{25},
  \bibinfo{number}{3} (\bibinfo{year}{2023}), \bibinfo{pages}{9--13}.
\newblock


\bibitem[Leonard(2023)]%
        {Leonard2023}
\bibfield{author}{\bibinfo{person}{Andrew Leonard}.}
  \bibinfo{year}{2023}\natexlab{}.
\newblock \bibinfo{title}{‘Dr. Google’ meets its match: Dr. ChatGPT}.
\newblock
\newblock
\urldef\tempurl%
\url{https://www.latimes.com/science/story/2023-09-08/dr-google-meets-its-match-dr-chatgpt}
\showURL{%
\tempurl}
\newblock
\shownote{Accessed: 09/11/2023}.


\bibitem[Li et~al\mbox{.}(2023)]%
        {li2023multi}
\bibfield{author}{\bibinfo{person}{Haoran Li}, \bibinfo{person}{Dadi Guo},
  \bibinfo{person}{Wei Fan}, \bibinfo{person}{Mingshi Xu}, \bibinfo{person}{Jie
  Huang}, \bibinfo{person}{Fanpu Meng}, {and} \bibinfo{person}{Yangqiu Song}.}
  \bibinfo{year}{2023}\natexlab{}.
\newblock \showarticletitle{Multi-step Jailbreaking Privacy Attacks on
  ChatGPT}. In \bibinfo{booktitle}{\emph{Findings of the Association for
  Computational Linguistics: EMNLP 2023}}. \bibinfo{publisher}{Association for
  Computational Linguistics}.
\newblock
\urldef\tempurl%
\url{https://doi.org/10.18653/v1/2023.findings-emnlp.272}
\showDOI{\tempurl}


\bibitem[Li et~al\mbox{.}(2021)]%
        {li2021large}
\bibfield{author}{\bibinfo{person}{Xuechen Li}, \bibinfo{person}{Florian
  Tramer}, \bibinfo{person}{Percy Liang}, {and} \bibinfo{person}{Tatsunori
  Hashimoto}.} \bibinfo{year}{2021}\natexlab{}.
\newblock \showarticletitle{Large language models can be strong differentially
  private learners}.
\newblock \bibinfo{journal}{\emph{arXiv preprint arXiv:2110.05679}}
  (\bibinfo{year}{2021}).
\newblock


\bibitem[Liang et~al\mbox{.}(2017)]%
        {liang2017privacy}
\bibfield{author}{\bibinfo{person}{Hai Liang}, \bibinfo{person}{Fei Shen},
  {and} \bibinfo{person}{King-wa Fu}.} \bibinfo{year}{2017}\natexlab{}.
\newblock \showarticletitle{Privacy protection and self-disclosure across
  societies: A study of global Twitter users}.
\newblock \bibinfo{journal}{\emph{new media \& society}} \bibinfo{volume}{19},
  \bibinfo{number}{9} (\bibinfo{year}{2017}), \bibinfo{pages}{1476--1497}.
\newblock


\bibitem[Liao and Vaughan(2023)]%
        {liao2023ai}
\bibfield{author}{\bibinfo{person}{Q~Vera Liao} {and}
  \bibinfo{person}{Jennifer~Wortman Vaughan}.} \bibinfo{year}{2023}\natexlab{}.
\newblock \showarticletitle{AI Transparency in the Age of LLMs: A
  Human-Centered Research Roadmap}.
\newblock \bibinfo{journal}{\emph{arXiv preprint arXiv:2306.01941}}
  (\bibinfo{year}{2023}).
\newblock


\bibitem[Lison et~al\mbox{.}(2021)]%
        {lison2021anonymisation}
\bibfield{author}{\bibinfo{person}{Pierre Lison}, \bibinfo{person}{Ildikó
  Pilán}, \bibinfo{person}{David Sanchez}, \bibinfo{person}{Montserrat Batet},
  {and} \bibinfo{person}{Lilja Øvrelid}.} \bibinfo{year}{2021}\natexlab{}.
\newblock \showarticletitle{Anonymisation Models for Text Data: State of the
  art, Challenges and Future Directions}. In
  \bibinfo{booktitle}{\emph{Proceedings of the 59th Annual Meeting of the
  Association for Computational Linguistics and the 11th International Joint
  Conference on Natural Language Processing (Volume 1: Long Papers)}}.
  \bibinfo{publisher}{Association for Computational Linguistics}.
\newblock
\urldef\tempurl%
\url{https://doi.org/10.18653/v1/2021.acl-long.323}
\showDOI{\tempurl}


\bibitem[Mai et~al\mbox{.}(2020)]%
        {mai2020user}
\bibfield{author}{\bibinfo{person}{Alexandra Mai}, \bibinfo{person}{Katharina
  Pfeffer}, \bibinfo{person}{Matthias Gusenbauer}, \bibinfo{person}{Edgar
  Weippl}, {and} \bibinfo{person}{Katharina Krombholz}.}
  \bibinfo{year}{2020}\natexlab{}.
\newblock \showarticletitle{User mental models of cryptocurrency systems-a
  grounded theory approach}. In \bibinfo{booktitle}{\emph{Sixteenth Symposium
  on Usable Privacy and Security (SOUPS 2020)}}. \bibinfo{pages}{341--358}.
\newblock


\bibitem[Majmudar et~al\mbox{.}(2022)]%
        {majmudar2022differentially}
\bibfield{author}{\bibinfo{person}{Jimit Majmudar}, \bibinfo{person}{Christophe
  Dupuy}, \bibinfo{person}{Charith Peris}, \bibinfo{person}{Sami Smaili},
  \bibinfo{person}{Rahul Gupta}, {and} \bibinfo{person}{Richard Zemel}.}
  \bibinfo{year}{2022}\natexlab{}.
\newblock \showarticletitle{Differentially private decoding in large language
  models}.
\newblock \bibinfo{journal}{\emph{arXiv preprint arXiv:2205.13621}}
  (\bibinfo{year}{2022}).
\newblock


\bibitem[McDonald et~al\mbox{.}(2019)]%
        {mcdonald2019reliability}
\bibfield{author}{\bibinfo{person}{Nora McDonald}, \bibinfo{person}{Sarita
  Schoenebeck}, {and} \bibinfo{person}{Andrea Forte}.}
  \bibinfo{year}{2019}\natexlab{}.
\newblock \showarticletitle{Reliability and inter-rater reliability in
  qualitative research: Norms and guidelines for CSCW and HCI practice}.
\newblock \bibinfo{journal}{\emph{Proceedings of the ACM on human-computer
  interaction}} \bibinfo{volume}{3}, \bibinfo{number}{CSCW}
  (\bibinfo{year}{2019}), \bibinfo{pages}{1--23}.
\newblock


\bibitem[McKee et~al\mbox{.}(2021)]%
        {mckee2021understanding}
\bibfield{author}{\bibinfo{person}{KR McKee}, \bibinfo{person}{X Bai}, {and}
  \bibinfo{person}{S Fiske}.} \bibinfo{year}{2021}\natexlab{}.
\newblock \showarticletitle{Understanding human impressions of artificial
  intelligence}.
\newblock \bibinfo{journal}{\emph{Preprint]. PsyArXiv. https://doi.
  org/10.31234/osf. io/5ursp}} (\bibinfo{year}{2021}).
\newblock


\bibitem[Mu et~al\mbox{.}(2023)]%
        {mu2023embodiedgpt}
\bibfield{author}{\bibinfo{person}{Yao Mu}, \bibinfo{person}{Qinglong Zhang},
  \bibinfo{person}{Mengkang Hu}, \bibinfo{person}{Wenhai Wang},
  \bibinfo{person}{Mingyu Ding}, \bibinfo{person}{Jun Jin},
  \bibinfo{person}{Bin Wang}, \bibinfo{person}{Jifeng Dai}, \bibinfo{person}{Yu
  Qiao}, {and} \bibinfo{person}{Ping Luo}.} \bibinfo{year}{2023}\natexlab{}.
\newblock \showarticletitle{EmbodiedGPT: Vision-Language Pre-Training via
  Embodied Chain of Thought}.
\newblock \bibinfo{journal}{\emph{arXiv preprint arXiv:2305.15021}}
  (\bibinfo{year}{2023}).
\newblock


\bibitem[Natarajan and Gombolay(2020)]%
        {natarajan2020effects}
\bibfield{author}{\bibinfo{person}{Manisha Natarajan} {and}
  \bibinfo{person}{Matthew Gombolay}.} \bibinfo{year}{2020}\natexlab{}.
\newblock \showarticletitle{Effects of anthropomorphism and accountability on
  trust in human robot interaction}. In \bibinfo{booktitle}{\emph{Proceedings
  of the 2020 ACM/IEEE international conference on human-robot interaction}}.
  \bibinfo{pages}{33--42}.
\newblock


\bibitem[Nissenbaum(2004)]%
        {nissenbaum2004privacy}
\bibfield{author}{\bibinfo{person}{Helen Nissenbaum}.}
  \bibinfo{year}{2004}\natexlab{}.
\newblock \showarticletitle{Privacy as contextual integrity}.
\newblock \bibinfo{journal}{\emph{Wash. L. Rev.}}  \bibinfo{volume}{79}
  (\bibinfo{year}{2004}), \bibinfo{pages}{119}.
\newblock


\bibitem[Nissenbaum(2020)]%
        {nissenbaum2020privacy}
\bibfield{author}{\bibinfo{person}{Helen Nissenbaum}.}
  \bibinfo{year}{2020}\natexlab{}.
\newblock \bibinfo{booktitle}{\emph{Privacy in context: Technology, policy, and
  the integrity of social life}}.
\newblock \bibinfo{publisher}{Stanford University Press}.
\newblock


\bibitem[Norman(2014)]%
        {norman2014some}
\bibfield{author}{\bibinfo{person}{Donald~A Norman}.}
  \bibinfo{year}{2014}\natexlab{}.
\newblock \showarticletitle{Some observations on mental models}.
\newblock In \bibinfo{booktitle}{\emph{Mental models}}.
  \bibinfo{publisher}{Psychology Press}, \bibinfo{pages}{15--22}.
\newblock


\bibitem[Ouyang et~al\mbox{.}(2022)]%
        {ouyang2022training}
\bibfield{author}{\bibinfo{person}{Long Ouyang}, \bibinfo{person}{Jeffrey Wu},
  \bibinfo{person}{Xu Jiang}, \bibinfo{person}{Diogo Almeida},
  \bibinfo{person}{Carroll Wainwright}, \bibinfo{person}{Pamela Mishkin},
  \bibinfo{person}{Chong Zhang}, \bibinfo{person}{Sandhini Agarwal},
  \bibinfo{person}{Katarina Slama}, \bibinfo{person}{Alex Ray},
  {et~al\mbox{.}}} \bibinfo{year}{2022}\natexlab{}.
\newblock \showarticletitle{Training language models to follow instructions
  with human feedback}.
\newblock \bibinfo{journal}{\emph{Advances in Neural Information Processing
  Systems}}  \bibinfo{volume}{35} (\bibinfo{year}{2022}),
  \bibinfo{pages}{27730--27744}.
\newblock


\bibitem[Pahune and Chandrasekharan(2023)]%
        {pahune2023several}
\bibfield{author}{\bibinfo{person}{Saurabh Pahune} {and} \bibinfo{person}{Manoj
  Chandrasekharan}.} \bibinfo{year}{2023}\natexlab{}.
\newblock \showarticletitle{Several categories of Large Language Models (LLMs):
  A Short Survey}.
\newblock \bibinfo{journal}{\emph{arXiv preprint arXiv:2307.10188}}
  (\bibinfo{year}{2023}).
\newblock


\bibitem[Pan et~al\mbox{.}(2020)]%
        {pan2020privacy}
\bibfield{author}{\bibinfo{person}{Xudong Pan}, \bibinfo{person}{Mi Zhang},
  \bibinfo{person}{Shouling Ji}, {and} \bibinfo{person}{Min Yang}.}
  \bibinfo{year}{2020}\natexlab{}.
\newblock \showarticletitle{Privacy risks of general-purpose language models}.
  In \bibinfo{booktitle}{\emph{2020 IEEE Symposium on Security and Privacy
  (SP)}}. IEEE, \bibinfo{pages}{1314--1331}.
\newblock


\bibitem[Peris et~al\mbox{.}(2023)]%
        {Peris_Dupuy_Majmudar_Parikh_Smaili_Zemel_Gupta_202}
\bibfield{author}{\bibinfo{person}{Charith Peris}, \bibinfo{person}{Christophe
  Dupuy}, \bibinfo{person}{Jimit Majmudar}, \bibinfo{person}{Rahil Parikh},
  \bibinfo{person}{Sami Smaili}, \bibinfo{person}{Richard Zemel}, {and}
  \bibinfo{person}{Rahul Gupta}.} \bibinfo{year}{2023}\natexlab{}.
\newblock \bibinfo{title}{Privacy in the Time of Language Models}.
\newblock
\newblock
\urldef\tempurl%
\url{https://doi.org/10.1145/3539597.3575792}
\showDOI{\tempurl}


\bibitem[Renaud et~al\mbox{.}(2014)]%
        {renaud2014doesn}
\bibfield{author}{\bibinfo{person}{Karen Renaud}, \bibinfo{person}{Melanie
  Volkamer}, {and} \bibinfo{person}{Arne Renkema-Padmos}.}
  \bibinfo{year}{2014}\natexlab{}.
\newblock \bibinfo{booktitle}{\emph{Why Doesn’t Jane Protect Her Privacy?}}
\newblock \bibinfo{publisher}{Springer International Publishing},
  \bibinfo{pages}{244–262}.
\newblock
\showISBNx{9783319085067}
\showISSN{1611-3349}
\urldef\tempurl%
\url{https://doi.org/10.1007/978-3-319-08506-7_13}
\showDOI{\tempurl}


\bibitem[Rutjes et~al\mbox{.}(2019)]%
        {rutjes2019considerations}
\bibfield{author}{\bibinfo{person}{Heleen Rutjes}, \bibinfo{person}{Martijn
  Willemsen}, {and} \bibinfo{person}{Wijnand IJsselsteijn}.}
  \bibinfo{year}{2019}\natexlab{}.
\newblock \showarticletitle{Considerations on explainable AI and users’
  mental models}. In \bibinfo{booktitle}{\emph{CHI 2019 Workshop: Where is the
  Human? Bridging the Gap Between AI and HCI}}. Association for Computing
  Machinery, Inc.
\newblock


\bibitem[Salda{\~n}a(2015)]%
        {saldana2015coding}
\bibfield{author}{\bibinfo{person}{Johnny Salda{\~n}a}.}
  \bibinfo{year}{2015}\natexlab{}.
\newblock \bibinfo{booktitle}{\emph{The coding manual for qualitative
  researchers}}.
\newblock \bibinfo{publisher}{Sage}.
\newblock


\bibitem[Shalaby et~al\mbox{.}(2020)]%
        {shalaby2020building}
\bibfield{author}{\bibinfo{person}{Walid Shalaby}, \bibinfo{person}{Adriano
  Arantes}, \bibinfo{person}{Teresa GonzalezDiaz}, {and}
  \bibinfo{person}{Chetan Gupta}.} \bibinfo{year}{2020}\natexlab{}.
\newblock \showarticletitle{Building chatbots from large scale domain-specific
  knowledge bases: Challenges and opportunities}. In
  \bibinfo{booktitle}{\emph{2020 IEEE International Conference on Prognostics
  and Health Management (ICPHM)}}. IEEE, \bibinfo{pages}{1--8}.
\newblock


\bibitem[Simmons et~al\mbox{.}(2011)]%
        {simmons2011believable}
\bibfield{author}{\bibinfo{person}{Reid Simmons}, \bibinfo{person}{Maxim
  Makatchev}, \bibinfo{person}{Rachel Kirby}, \bibinfo{person}{Min~Kyung Lee},
  \bibinfo{person}{Imran Fanaswala}, \bibinfo{person}{Brett Browning},
  \bibinfo{person}{Jodi Forlizzi}, {and} \bibinfo{person}{Majd Sakr}.}
  \bibinfo{year}{2011}\natexlab{}.
\newblock \showarticletitle{Believable robot characters}.
\newblock \bibinfo{journal}{\emph{AI Magazine}} \bibinfo{volume}{32},
  \bibinfo{number}{4} (\bibinfo{year}{2011}), \bibinfo{pages}{39--52}.
\newblock


\bibitem[Solove(2020)]%
        {solove2021myth}
\bibfield{author}{\bibinfo{person}{Daniel~J. Solove}.}
  \bibinfo{year}{2020}\natexlab{}.
\newblock \showarticletitle{The Myth of the Privacy Paradox}.
\newblock \bibinfo{journal}{\emph{SSRN Electronic Journal}}
  (\bibinfo{year}{2020}).
\newblock
\showISSN{1556-5068}
\urldef\tempurl%
\url{https://doi.org/10.2139/ssrn.3536265}
\showDOI{\tempurl}


\bibitem[Stutzman et~al\mbox{.}(2011)]%
        {stutzman2011factors}
\bibfield{author}{\bibinfo{person}{Fred Stutzman}, \bibinfo{person}{Robert
  Capra}, {and} \bibinfo{person}{Jamila Thompson}.}
  \bibinfo{year}{2011}\natexlab{}.
\newblock \showarticletitle{Factors mediating disclosure in social network
  sites}.
\newblock \bibinfo{journal}{\emph{Computers in Human Behavior}}
  \bibinfo{volume}{27}, \bibinfo{number}{1} (\bibinfo{year}{2011}),
  \bibinfo{pages}{590--598}.
\newblock


\bibitem[Taver(2023)]%
        {taver2023chatgpt}
\bibfield{author}{\bibinfo{person}{Mikhail Taver}.}
  \bibinfo{year}{2023}\natexlab{}.
\newblock \bibinfo{title}{ChatGPT is Coming to Finance, So Let’s Talk About
  the Risks and Rewards}.
\newblock
  \bibinfo{howpublished}{\url{https://www.unite.ai/chatgpt-is-coming-to-finance-so-lets-talk-about-the-risks-and-rewards/}}.
\newblock
\newblock
\shownote{Accessed: 09/11/2023}.


\bibitem[Ur et~al\mbox{.}(2015)]%
        {ur2015measuring}
\bibfield{author}{\bibinfo{person}{Blase Ur}, \bibinfo{person}{Sean~M Segreti},
  \bibinfo{person}{Lujo Bauer}, \bibinfo{person}{Nicolas Christin},
  \bibinfo{person}{Lorrie~Faith Cranor}, \bibinfo{person}{Saranga Komanduri},
  \bibinfo{person}{Darya Kurilova}, \bibinfo{person}{Michelle~L Mazurek},
  \bibinfo{person}{William Melicher}, {and} \bibinfo{person}{Richard Shay}.}
  \bibinfo{year}{2015}\natexlab{}.
\newblock \showarticletitle{Measuring $\{$Real-World$\}$ Accuracies and Biases
  in Modeling Password Guessability}. In \bibinfo{booktitle}{\emph{24th USENIX
  Security Symposium (USENIX Security 15)}}. \bibinfo{pages}{463--481}.
\newblock


\bibitem[Waldman(2018)]%
        {waldman2018privacy}
\bibfield{author}{\bibinfo{person}{Ari~Ezra Waldman}.}
  \bibinfo{year}{2018}\natexlab{}.
\newblock \bibinfo{booktitle}{\emph{Privacy as trust: Information privacy for
  an information age}}.
\newblock \bibinfo{publisher}{Cambridge University Press}.
\newblock


\bibitem[Waldman(2021)]%
        {waldman2021industry}
\bibfield{author}{\bibinfo{person}{Ari~Ezra Waldman}.}
  \bibinfo{year}{2021}\natexlab{}.
\newblock \bibinfo{booktitle}{\emph{Industry unbound: The inside story of
  privacy, data, and corporate power}}.
\newblock \bibinfo{publisher}{Cambridge University Press}.
\newblock


\bibitem[Wang et~al\mbox{.}(2011)]%
        {wang2011regretted}
\bibfield{author}{\bibinfo{person}{Yang Wang}, \bibinfo{person}{Gregory
  Norcie}, \bibinfo{person}{Saranga Komanduri}, \bibinfo{person}{Alessandro
  Acquisti}, \bibinfo{person}{Pedro~Giovanni Leon}, {and}
  \bibinfo{person}{Lorrie~Faith Cranor}.} \bibinfo{year}{2011}\natexlab{}.
\newblock \showarticletitle{" I regretted the minute I pressed share" a
  qualitative study of regrets on Facebook}. In
  \bibinfo{booktitle}{\emph{Proceedings of the seventh symposium on usable
  privacy and security}}. \bibinfo{pages}{1--16}.
\newblock


\bibitem[Weidinger et~al\mbox{.}(2021)]%
        {weidinger2021ethical}
\bibfield{author}{\bibinfo{person}{Laura Weidinger}, \bibinfo{person}{John
  Mellor}, \bibinfo{person}{Maribeth Rauh}, \bibinfo{person}{Conor Griffin},
  \bibinfo{person}{Jonathan Uesato}, \bibinfo{person}{Po-Sen Huang},
  \bibinfo{person}{Myra Cheng}, \bibinfo{person}{Mia Glaese},
  \bibinfo{person}{Borja Balle}, \bibinfo{person}{Atoosa Kasirzadeh},
  {et~al\mbox{.}}} \bibinfo{year}{2021}\natexlab{}.
\newblock \showarticletitle{Ethical and social risks of harm from language
  models}.
\newblock \bibinfo{journal}{\emph{arXiv preprint arXiv:2112.04359}}
  (\bibinfo{year}{2021}).
\newblock


\bibitem[Ye et~al\mbox{.}(2023)]%
        {ye2023improved}
\bibfield{author}{\bibinfo{person}{Yang Ye}, \bibinfo{person}{Hengxu You},
  {and} \bibinfo{person}{Jing Du}.} \bibinfo{year}{2023}\natexlab{}.
\newblock \showarticletitle{Improved Trust in Human-Robot Collaboration With
  ChatGPT}.
\newblock \bibinfo{journal}{\emph{IEEE Access}}  \bibinfo{volume}{11}
  (\bibinfo{year}{2023}).
\newblock
\showISSN{2169-3536}
\urldef\tempurl%
\url{https://doi.org/10.1109/access.2023.3282111}
\showDOI{\tempurl}


\bibitem[Yu et~al\mbox{.}(2021)]%
        {yu2021differentially}
\bibfield{author}{\bibinfo{person}{Da Yu}, \bibinfo{person}{Saurabh Naik},
  \bibinfo{person}{Arturs Backurs}, \bibinfo{person}{Sivakanth Gopi},
  \bibinfo{person}{Huseyin~A Inan}, \bibinfo{person}{Gautam Kamath},
  \bibinfo{person}{Janardhan Kulkarni}, \bibinfo{person}{Yin~Tat Lee},
  \bibinfo{person}{Andre Manoel}, \bibinfo{person}{Lukas Wutschitz},
  {et~al\mbox{.}}} \bibinfo{year}{2021}\natexlab{}.
\newblock \showarticletitle{Differentially private fine-tuning of language
  models}.
\newblock \bibinfo{journal}{\emph{arXiv preprint arXiv:2110.06500}}
  (\bibinfo{year}{2021}).
\newblock


\bibitem[Zeng et~al\mbox{.}(2017)]%
        {zeng2017end}
\bibfield{author}{\bibinfo{person}{Eric Zeng}, \bibinfo{person}{Shrirang Mare},
  {and} \bibinfo{person}{Franziska Roesner}.} \bibinfo{year}{2017}\natexlab{}.
\newblock \showarticletitle{End user security and privacy concerns with smart
  homes}. In \bibinfo{booktitle}{\emph{thirteenth symposium on usable privacy
  and security (SOUPS 2017)}}. \bibinfo{pages}{65--80}.
\newblock


\bibitem[Zhang et~al\mbox{.}(2021)]%
        {zhang2021counterfactual}
\bibfield{author}{\bibinfo{person}{Chiyuan Zhang}, \bibinfo{person}{Daphne
  Ippolito}, \bibinfo{person}{Katherine Lee}, \bibinfo{person}{Matthew
  Jagielski}, \bibinfo{person}{Florian Tram{\`e}r}, {and}
  \bibinfo{person}{Nicholas Carlini}.} \bibinfo{year}{2021}\natexlab{}.
\newblock \showarticletitle{Counterfactual memorization in neural language
  models}.
\newblock \bibinfo{journal}{\emph{arXiv preprint arXiv:2112.12938}}
  (\bibinfo{year}{2021}).
\newblock


\bibitem[Zhang et~al\mbox{.}(2023)]%
        {zhang2023huatuogpt}
\bibfield{author}{\bibinfo{person}{Hongbo Zhang}, \bibinfo{person}{Junying
  Chen}, \bibinfo{person}{Feng Jiang}, \bibinfo{person}{Fei Yu},
  \bibinfo{person}{Zhihong Chen}, \bibinfo{person}{Guiming Chen},
  \bibinfo{person}{Jianquan Li}, \bibinfo{person}{Xiangbo Wu},
  \bibinfo{person}{Zhang Zhiyi}, \bibinfo{person}{Qingying Xiao},
  \bibinfo{person}{Xiang Wan}, \bibinfo{person}{Benyou Wang}, {and}
  \bibinfo{person}{Haizhou Li}.} \bibinfo{year}{2023}\natexlab{}.
\newblock \showarticletitle{HuatuoGPT, Towards Taming Language Model to Be a
  Doctor}. In \bibinfo{booktitle}{\emph{Findings of the Association for
  Computational Linguistics: EMNLP 2023}}. \bibinfo{publisher}{Association for
  Computational Linguistics}.
\newblock
\urldef\tempurl%
\url{https://doi.org/10.18653/v1/2023.findings-emnlp.725}
\showDOI{\tempurl}


\bibitem[Zhao et~al\mbox{.}(2023)]%
        {zhao2023survey}
\bibfield{author}{\bibinfo{person}{Wayne~Xin Zhao}, \bibinfo{person}{Kun Zhou},
  \bibinfo{person}{Junyi Li}, \bibinfo{person}{Tianyi Tang},
  \bibinfo{person}{Xiaolei Wang}, \bibinfo{person}{Yupeng Hou},
  \bibinfo{person}{Yingqian Min}, \bibinfo{person}{Beichen Zhang},
  \bibinfo{person}{Junjie Zhang}, \bibinfo{person}{Zican Dong},
  {et~al\mbox{.}}} \bibinfo{year}{2023}\natexlab{}.
\newblock \showarticletitle{A survey of large language models}.
\newblock \bibinfo{journal}{\emph{arXiv preprint arXiv:2303.18223}}
  (\bibinfo{year}{2023}).
\newblock


\bibitem[Zheng et~al\mbox{.}(2023)]%
        {zheng2023judging}
\bibfield{author}{\bibinfo{person}{Lianmin Zheng}, \bibinfo{person}{Wei-Lin
  Chiang}, \bibinfo{person}{Ying Sheng}, \bibinfo{person}{Siyuan Zhuang},
  \bibinfo{person}{Zhanghao Wu}, \bibinfo{person}{Yonghao Zhuang},
  \bibinfo{person}{Zi Lin}, \bibinfo{person}{Zhuohan Li},
  \bibinfo{person}{Dacheng Li}, \bibinfo{person}{Eric Xing}, {et~al\mbox{.}}}
  \bibinfo{year}{2023}\natexlab{}.
\newblock \showarticletitle{Judging LLM-as-a-judge with MT-Bench and Chatbot
  Arena}.
\newblock \bibinfo{journal}{\emph{arXiv preprint arXiv:2306.05685}}
  (\bibinfo{year}{2023}).
\newblock


\bibitem[Zlatolas et~al\mbox{.}(2015)]%
        {zlatolas2015privacy}
\bibfield{author}{\bibinfo{person}{Lili~Nemec Zlatolas},
  \bibinfo{person}{Tatjana Welzer}, \bibinfo{person}{Marjan Heri{\v{c}}ko},
  {and} \bibinfo{person}{Marko H{\"o}lbl}.} \bibinfo{year}{2015}\natexlab{}.
\newblock \showarticletitle{Privacy antecedents for SNS self-disclosure: The
  case of Facebook}.
\newblock \bibinfo{journal}{\emph{Computers in Human Behavior}}
  \bibinfo{volume}{45} (\bibinfo{year}{2015}), \bibinfo{pages}{158--167}.
\newblock


\bibitem[Z{\l}otowski et~al\mbox{.}(2015)]%
        {zlotowski2015anthropomorphism}
\bibfield{author}{\bibinfo{person}{Jakub Z{\l}otowski}, \bibinfo{person}{Diane
  Proudfoot}, \bibinfo{person}{Kumar Yogeeswaran}, {and}
  \bibinfo{person}{Christoph Bartneck}.} \bibinfo{year}{2015}\natexlab{}.
\newblock \showarticletitle{Anthropomorphism: opportunities and challenges in
  human--robot interaction}.
\newblock \bibinfo{journal}{\emph{International journal of social robotics}}
  \bibinfo{volume}{7} (\bibinfo{year}{2015}), \bibinfo{pages}{347--360}.
\newblock


\bibitem[Zuboff(2023)]%
        {zuboff2023age}
\bibfield{author}{\bibinfo{person}{Shoshana Zuboff}.}
  \bibinfo{year}{2023}\natexlab{}.
\newblock \showarticletitle{The age of surveillance capitalism}.
\newblock In \bibinfo{booktitle}{\emph{Social Theory Re-Wired}}.
  \bibinfo{publisher}{Routledge}, \bibinfo{pages}{203--213}.
\newblock


\end{thebibliography}


\appendix

\section{PII coding criteria}
\label{sec:pii-coding-criteria}
\subsection{Rules for Determining Non PII vs. PII} 
If the data type is incorrect (e.g., it is not even a person’s name at all), or it’s public information (e.g., public figures, or public information online), select NA, meaning it is a non PII type.

If it’s not clear from the text whether it is a random name or a real person’s name, we can generalize the case to similar scenarios that we may encounter in real life, and think about whether it makes sense to include a real person’s name in that situation, and if so, label it as a PII type. A random name being used might be if the name appears to be used as a placeholder for a real person’s name. For example, instances of the name, “John Doe”, being used are likely using it as a placeholder.

If unsure whether certain information is public or private, attempt to search it online. If the information is easily found online and is not linked to a specific individual, then it is considered more public information (non PII). If the information is only available to individuals who directly interact with the owner, or the information is linked directly to a specific person that can be identified, then it is considered more private information. If the information may be used to infer some personal information of the user or some other people, but it’s not clear whether it is directly associated with these people, don’t label them as PII.

\subsection{Rules for Determining Actor Type (Self/Others/Both/Unknown)}

If the text doesn’t provide sufficient evidence for us to determine the relationship between the user and the persons mentioned in the text, label as Unknown; otherwise, label as Self, Others, or Both according to the context. 

A label of Self means that the PII data appears to be about the human who conversed with ChatGPT. If the PII data appears to be about other people, then label as Others. If the PII data appears to be of both cases, where it is related to the ChatGPT user and other people, label as Both. Otherwise, if there is limited information, or it is uncertain who the PII data belongs to, then label as Unknown.

If this text explicitly mentions the PII in the context of belonging to the user, such as saying “my information” or “I am xxx”, label as Self. Sometimes if the tone or other information imply that certain topics are associated with a specific person, then label it as PII. For example, if a user asked ChatGPT to perform a very specific task involving a URL, it may be safe to assume that the URL is associated with the user themself. 

If there are multiple categories that may apply to the PII we will select Unknown for now, but make a note of it. This may occur in the case of the user presenting multiple pieces of PII, in which case, it may be true that the PII belongs to third party individuals, or it belongs to both the user themself and third party individuals.

\subsection{Rules About PIIs not Detected by the Algorithm}
There may be cases of non-detected PIIs by the algorithm. In this case, if we notice PII that wasn’t detected, we should still label them, and make a note about them. 

\subsection{Rules about Specific Categories of PII}
When labeling the category of PERSON, first names, last names, full names, and aliases all count as being this PII type. 

For DATE\_TIME, there may be instances of the current date being included in web search results, in which case we label as Self. Timestamps in messages between individuals should be labeled as Both. Cases of transaction histories, or other examples of seemingly personal documents, should be labeled as Self.

For the IP\_ADDRESS category, private IP addresses are not considered PII. 

In the case of URL, if the URL helps identify a specific person, e.g., the URL is the website of the person’s company, label it as PII. If the URL contains tracking parameters such as \_ga, \_gac, label as Unknown.

For NRP and LOCATION, it may be more difficult to determine if the data is considered PII. Thus, it is important to consider the context, especially the prompt written by the user. If, given the context, we infer that the detected data describes the NRP or location of a particular person in a relatively private context, label it as PII.

\section{Codebook for Dataset Analysis}
\label{sec:codebook-dataset-scenarios}
Our codebook of the ShareGPT dataset analysis results is shown in \autoref{tab:codebook-dataset-scenarios}.

\begin{table*}[]
    \centering
    \caption{Codebook for Dataset Analysis}
    \begin{tabular}{p{0.06\linewidth} p{0.1\linewidth} p{0.8\linewidth}}
    \toprule
    Theme& Label& Definition\\
    \midrule
    Context& Work-Related& Scenarios related to the workplace/industry context. 
    
    \textbf{Potential Privacy Risks}: There could be potential privacy risks with sharing confidential business information, marketing ideas and strategies, workplace communications, and many other industry related information. It could create privacy risks for the user sharing, since it might be considered a violation of workplace etiquette or rules by sharing industry related information.\\
    
    & Academic-Related& Scenarios related to an academic context.
    
    \textbf{Potential Privacy Risks}: There could be potential privacy risks with asking ChatGPT to generate academic related texts, finish assignments, or complete tasks related to academic scenarios. There might be risks of plagiarism, thus leading to reputational harm if it was discovered that these users used AI to generate texts that they claim to be their own.\\

    & Life-Related& Scenarios related to an individual’s personal life, emotions, problems, or more.
    
    \textbf{Potential Privacy Risks}: There could be potential privacy risks with sharing personal information and experiences with ChatGPT. It could potentially be leaked, thus leading to reputational harm or relationship harm if the information involves third party individuals. Users might not want their sensitive information to be shared to the public. The information being leaked or used by other organizations could also lead to the data being used for targeting by advertisers.\\
    Topic& Business& Scenarios related to business, spanning from generating and ideating business ideas, to asking for business advice. 
    
    \textbf{Potential Privacy Risks}: There are potential privacy risks with sharing business related details, such as ideas, plans, strategies, etc… Some of that information might be confidential, or it could imply more information about the user, such as their location or company name.\\
    
    & Assignment& Scenarios where the user is asking ChatGPT assignment-related questions, or they are asking for assistance in finishing tasks for assignments.
    
    \textbf{Potential Privacy Risks}: There could be potential privacy risks with asking ChatGPT to help on assignment, such as risk of plagiarism accusations, and more. Also, the assignment details might imply details about the user, such as their academic direction.\\

    & Programming& Scenarios related to programming and coding. The scenarios span anywhere from code inquiries to code generation to debugging help.
    
    \textbf{Potential Privacy Risks}: Usually in these scenarios, there are limited privacy risks that exist. However, in some cases, the user will include personal information, such as phone numbers or email addresses, in the shared code.\\

    & Financial& Scenarios involving financial cases and inquiries.
    
    \textbf{Potential Privacy Risks}: There might be risks with sharing financial information, such as transaction histories, or other financial details, such as an increased chance of fraudulent activity with the financial accounts.\\

    & Legal& Scenarios related to law of legal cases. This can include the user sharing legal cases or seeking legal advice.
    
    \textbf{Potential Privacy Risks}: There might be potential privacy risks involving multiple parties, such as the attorney, clients, or the legal case in general. There could be consequences with sharing confidential information about a case with ChatGPT, in the case it got leaked.\\

    & Medical& Scenarios that include medical related inquiries, tasks, or other similar conversations with ChatGPT.
    
    \textbf{Potential Privacy Risks}: There could be potential privacy risks leading to autonomy harms if the people involved do not know their medical information is being shared to ChatGPT.\\

    & Life& Scenarios related to daily life, personal inquiries about life/relationships, and more.
    
    \textbf{Potential Privacy Risks}: There could be potential privacy risks with sharing personal information about the user’s life, as well as experiences with ChatGPT. If it were potentially leaked, it could lead to reputational harm or relationship harm if the information involves third party individuals.\\

    & Entertainment& Scenarios involving entertainment-related topics.
    
    \textbf{Potential Privacy Risks}: Most scenarios for this topic have limited privacy risks since the content shared or generated is mostly fictional. However, when the user begins incorporating personal details into the conversation, there could be privacy risks involved. Also, depending on the type of scenario, there could be reputational harm for the user if they ask ChatGPT to generate some more inappropriate content.\\

    Purpose& Generate content/ writing& Generating different forms of written or text-based content. 
    
    \textbf{Potential Privacy Risks}: There might be privacy risks leading to reputational harm if it was made known that the content/writing was generated by AI. The user could potentially be accused of plagiarism if they tried using the generated content publicly.\\
\bottomrule
    \end{tabular}
    \label{tab:codebook-dataset-scenarios}
\end{table*}

\begin{table*}[]
    \centering
    \caption*{\autoref{tab:codebook-dataset-scenarios} (continued)}
    \begin{tabular}{p{0.06\linewidth} p{0.1\linewidth} p{0.8\linewidth}}
    \toprule
    Theme& Label& Definition\\
    \midrule

    & Generating plans/advice& Generating solutions or advice in response to a problem or query.
    
    \textbf{Potential Privacy Risks}: There might be privacy risks with the user sharing personal details in order for ChatGPT to give advice or generate plans. The user might share more information in order to provide enough context for ChatGPT to generate a useful plan or give good advice.\\

    & Answering questions& Using ChatGPT to answer direct questions that might typically be answered via a web search, or interpreting/responding to user input.
    
    \textbf{Potential Privacy Risks}: Privacy risks might occur less in this circumstance, since the user typically shares details that would be included in a Google search as well. However, the information could still be used to imply details about the user, such as personal interests or similar things.\\

    & Data analysis& Using ChatGPT to analyze content input by user.
    
    \textbf{Potential Privacy Risks}: There could be risks of data leakage or privacy violations by sharing the data needed for analysis. A lot of times, this could also lead to discovery of more information about the user’s work if the data shared is about work.\\

    & Casual conversations& Users engage in casual, objective-less chatting with ChatGPT. The aim is the act of conversing itself, not to reach a specific goal.
    
    \textbf{Potential Privacy Risks}: There could be a lot of privacy risks since the user might feel more comfortable with talking to ChatGPT like another human. They might start revealing more personal details about their life, rather than the typical identifying information like an address or phone number. There might also be instances of the shared details from the conversation being used to identity the user.\\

    Way to prompt& Direct command& Users ask straightforward questions without providing much context or background information. It typically involves limited data sharing because users input minimum information required to formulate their questions. 
    
    \textbf{Potential Privacy Risks}: While there might be limited data sharing since user’s likely only include the most necessary details in their questions/commands, there is still risk of them sharing information that could imply personal interests.\\

    & Interactively define the tasks& Users engage in multi-round interactions or chats with ChatGPT. The nature of these ongoing conversations can sometimes lead to users sharing more data with ChatGPT, especially as the conversation deepens or becomes more complex.
    
    \textbf{Potential Privacy Risks}: There are potential risks of users sharing more and more information with ChatGPT. For example, if the user does not completely fulfill their tasks with the original information provided to ChatGPT, they might feel inclined to provide more specific information, which may lead to sharing more sensitive data about themselves. \\

    & Handle the tasks based on given text& Users provide more extensive background information before asking for a response or action from ChatGPT. This approach can lead to more significant data sharing because users are disclosing more context for ChatGPT to process.
    
    \textbf{Potential Privacy Risks}: There could be potential privacy risks both with the shared text and with the tasks involved. The shared text might present information that could be used to imply new pieces of information about the user, such as their interests, occupation, or location. Also, depending on the tasks given to ChatGPT, there could also be risks involving plagiarism or reputational harm if the user asked ChatGPT to write something for them.\\

    & Role-playing& Users assign a specific role to ChatGPT and provide related content to request ChatGPT to complete tasks or generate responses. Although this method may involve sharing a lot of information, the data’s authenticity and ownership might be ambiguous because the information is framed within a role-play scenario.
    
    \textbf{Potential Privacy Risks}: In these circumstances, sometimes, the user-assigned role to ChatGPT might imply information about the user themself, such as their own occupation or feelings toward a topic. This information might be used to trace back to the user and discover more information about their identity. \\

    & Jailbreaking& This is a special category where users attempt to push beyond the designed limits of ChatGPT. This can lead to unexpected outputs, sometimes even generating potentially harmful or violent content. While this does not necessarily imply higher levels of data sharing, the behavior to generate unexpected output might present a potential risk.
    
    \textbf{Potential Privacy Risks}: There could be potential privacy risks leading to reputational harm if the conversations got leaked. Since in these circumstances, the user is usually trying to violate some community guidelines, or are inquiring about potentially harmful or violent actions, it could harm the user’s reputation. \\

    \bottomrule
    \end{tabular}
\end{table*}

\section{Interview Script}
\label{sec:interview-script}
\subsection{Introduction}
Hello! Nice to meet you. So happy you could join us today. I'm a researcher at Northeastern University and looking forward to chatting with you. 

Before we start, I will introduce a bit about the interview, so you can know more about this. For the interview, we'd like to know the challenges that GPT users have encountered when they are dealing with tasks that require them to share personal data with GPT, and their questions and confusions about how ChatGPT uses their data. The goal of our research is to gain a better understanding of users’ perspectives and challenges so that we can design systems like ChatGPT that are safer and more respectful.

Feel free to answer only the questions you're comfortable with. If there’s a question that you don’t want to discuss, feel free to let me know. It won’t affect your compensation.

We will record the chat but rest assured, all your responses are only accessible to researchers in our team and will be kept confidential. Does that work for you?

Can I start recording the session?

(After getting their affirmative answer) Great, I will start to record now. Let's get started!

\subsection{Basic Use Experience}
First, let’s talk about your experience with ChatGPT. 
\begin{itemize}
    \item When did you start using it? Why did you start using it?
    \item What do you usually use it for? 
\end{itemize}

Interesting, now I’d like you to show me 3 examples of conversation history including personal data you have prepared. 
Before that, just to confirm, have you blurred or covered any information that you do not want others to see in these 3 examples?
\begin{itemize}
    \item \emph{If not:} 
    \begin{itemize}
        \item Don't worry, you can take some time to do it now. 
        [Provide URL with instructions and examples of ChatGPT conversations with blurred information]
    \end{itemize}
\end{itemize}
That’s great! Feel free to share your screen and show it to us whenever you're ready.
[Participant shares his/her screen]
\begin{itemize}
    \item \emph{If some use cases are mentioned in the pre-screening survey, but not covered in the examples:} 
    \begin{itemize}
        \item I have noticed your responses in the questionnaire said you have used ChatGPT for [specific scenarios mentioend in the pre-screening surveys].
        \item Could you explain how you do it in this context? 
        \item Could you pick an example to describe what information you have input to ChatGPT to get responses in this context? You don't need to show the conversation to us.
    \end{itemize}
\end{itemize}

\subsection{Reflections: Privacy Concerns and Challenges in Preserving Privacy in Selected Conversation Examples}
\begin{itemize}
    \item Could you tell me the story about this chat? What’s your primary goal for this chat? 
    \item In this chat, What is the information you covered? 
    \item Why do you think this is the information you're not comfortable sharing with others?
    \item Have you had any concerns about sharing such data with ChatGPT? 
    \item Have you ever considered addressing these concerns?
    \item What methods have you tried or thought about? Give more details about that process?
    \item Did you encounter any challenges or difficulties when trying to do this? 
    \item \emph{If he/she has shared other people’s information:}
    \begin{itemize}
        \item Have you thought that other people might disclose your personal information as well? 
    \end{itemize}
    \item \emph{After discussion on all the examples:}
    \begin{itemize}
        \item Have you wanted to use ChatGPT to do something but refrained from doing so because of concerns about your privacy? Could you tell us more about it?
        \item Could you explain more about your thought process? 
        \item \emph{If we noticed a conflict between what they mentioned using ChatGPT for and their concerns:} 
        \begin{itemize}
            \item I’d like to give you a minute to review your chat histories that use xxx data.
In what circumstance, you would still do so even if you thought of that?
And under what circumstance, you would stop doing so based on the concern you mentioned?
        \end{itemize}
    \item Interesting. Let’s think about the privacy issue from the other side.
Have you used ChatGPT for any tasks to protect your privacy? 
Such as, you feel more comfortable asking ChatGPT to do certain tasks or sharing certain data with ChatGPT than other choices? 
    \end{itemize}    
\end{itemize}

\subsection{Mental Model on GPT-based Conversational Agent}
Now, let’s go to the next part. [Open up writeboard on Zoom]
\subsubsection{General System}
[For ChatGPT users]

In this session, I’d like to learn more about your understanding on how the system works and how the system will use your data. Please feel free to share your best understanding about how the system handles your data. Your input will be extremely useful for us to understand the limitations of current system design. And we won’t judge your answer and your compensation won’t be affected. Does it sound good?

Great, now we will do a drawing exercise. I’m going to ask you to explain your perceptions and ideas about how ChatGPT works — keeping in mind how things work “behind the scenes” — when you are chatting with ChatGPT. Imagine you ask ChatGPT a question, how does the data you input go through the system? 

You can use the drawing and texting tools on the left of the screen [show instructions], to draw how you think the ChatGPT works, what will happen to your input data after it is submitted through the interface [show the starting point]. Please talk aloud and explain your thought processes while you are drawing.

Assume that one year later, other users asked ChatGPT similar questions. Do you think that the information you provided a year ago could potentially influence the responses produced by ChatGPT? Why or why not?\\

\emph{If the participants showed misunderstandings, use the following part to debrief them on the process:}

[Ask to follow cursor in the Whiteboard]
\begin{itemize}
    \item  Great, you have mentioned most of the parts. I’d like to give you more information about how it works and how your data flows within ChatGPT. 
    \item After the data gets input through the interface, your input will be preprocessed first and then be sent to the AI model (we have GPT here) which is trained by large public internet data. The GPT will generate the responses based on your input and then return it to you.
    \item During the process of sending user input from the interface to the processing unit, your data will be uploaded to OpenAI's remote server. So the later process will all run on the remote server until the responses are returned.
   \item This is generally what happens behind when you're using this. 
\end{itemize}

\emph{[Data retention/ storage/storing in a database]:}

It is worth mentioning that your input data will be stored in the data storage which is remote as well. 
\begin{itemize}
    \item Are you aware that your data will be stored in the system before?
    \item \emph{If not:}
    \begin{itemize}
        \item Will this have any impact on your preferences of sharing data with ChatGPT?
    \end{itemize}
    \item \emph{If so:}
    \begin{itemize}
        \item Did it have any impact on your preferences of sharing data with ChatGPT?
    \end{itemize}
\end{itemize}

\emph{[Model training \& Memorization]:}

Another important process is model training. Parts of the data stored will be used for ChatGPT training. So the GPT AI model can improve overtime.
\begin{itemize}
    \item Are you aware of that before? How did you know that?
    \item What do you think of their use of your conversations for training models?
    \item Do you think there are any risks that may be caused by using your or other user conversations for model training?
\end{itemize}

There has been some research showing a memorization risk in the large language models. These models may memorize parts of the data used for training. That means, OpenAI’s future models may memorize some data that you provided, and because the same model is used by all the users, your information may be leaked to other people when they ask certain questions to the model.

\emph{[There was research that showed that GPT-3 offered detailed private information about the Editor-in-Chief of MIT Technology Review including his family members, work address, and phone number.]}
\begin{itemize}
    \item Are you aware of training or memorization?
    \item After learning about that, what do you think about your preferences for data retention?
    \item Are there any data types you are concerned about being memorized?
\end{itemize}

\subsubsection{Detailed Function}
Great, let’s dive into the detailed functions. 
Similar to the previous session, we’re evaluating the system design, not your capacity.
Feel free to share your best understanding about the questions that we are going to discuss. The correctness of your answers won’t affect your compensation.
We will have discussion towards the interface of ChatGPT. So, please open the ChatGPT interface and share your screen with me.
You can choose to only share certain parts of the interface, specifically, you may hide any conversation titles for your privacy. You can follow the instructions here if needed. 
[Guidance to share portion of the screen]

\emph{[Chat history]:}

We have talked about the Data Storage:
\begin{itemize}
    \item Have you done anything with the Chat history before? For what? (e.g. delete, export, sharing link) 
    \item What’s your expectation after deleting the Chat history? (Do you think it will be completely deleted from the storage?) 
\end{itemize}
According to OpenAI’s policy, the history will be retained for a maximum of 30 days after you delete it through the interface.
\begin{itemize}
    \item Did you know that before?
\end{itemize}

\emph{[Training data opt-out]:} 

[For ChatGPT users]

As we discussed before, OpenAI will use the user's conversations with ChatGPT to train their models. 
\begin{itemize}
    \item Do you know any method to opt-out and stop your data being used in model training? Have you thought of that before? 
\end{itemize}

Yes, there is an opt-out option there, the default setting is opt-in.
\begin{itemize}
    \item Did you know that before? Do you know how to opt out of data training?
    \item \emph{If they say yes:}
    \begin{itemize}
        \item How did you know that?
        \item Could you show me how to do it?
    \end{itemize}
    \item \emph{If they say no:}
    \begin{itemize}
        \item Maybe you can have a try to find the opt-out option.
    \end{itemize}
    \item \emph{If they fail:}
    \begin{itemize}
        \item You can first find out the “settings” button in the bottom left corner. Then click the ``Data controls'', you will see it in the first line.
        \item Now, you know this option. Would you consider opting out of data training? Why?
    \end{itemize}
\end{itemize}

\emph{[Account sharing]:}

\begin{itemize}
    \item Have you shared your ChatGPT account or OpenAI account with anyone else? Why or why not?
    \item Have you shared any other account (besides ChatGPT) with others? How similar or different do you perceive about these account sharing practices compared with ChatGPT account sharing?
\end{itemize}

\emph{[Sharing conversations with other users]:}

\begin{itemize}
    \item Have you considered sharing conversations with others? Could you tell us more about it? (When, why, how, shared content, with whom)
    \item Have you used any browser extension for sharing your conversations or the native sharing feature? Why or why not?
    \item \emph{If they have used them:}
    \begin{itemize}
        \item \emph{Do you remember, under what circumstances did you use the sharing tools?}
    \end{itemize}
    \item \emph{If they have used the ChatGPT native ``shared links'' feature:}
    \begin{itemize}
        \item Who do you think can see the content in this link?
        \item Do you know that anyone with the link can view and continue the linked conversation?
        \item Do you know how to delete or invalidate the link? Could you try to do so?
        \item \emph{If they fail:}
        \begin{itemize}
            \item You can first go to the “settings” in the bottom left corner. Then click the “Data controls”. In the 2nd line you can manage your shared links.
        \end{itemize}
    \end{itemize}
    \item \emph{If they have used (a) chrome extension(s):}
    \begin{itemize}
        \item What is it/are they?
        \item Who do you think can see the conversations you shared?
        \item Do you know about [Public dataset curated from leaked conversations like ShareGPT]? 
        \item \emph{If not, describe.}
        \item What do you think about the impact of this kind of data leak on your data?
        \item What type of your own data pops up in your mind that would be sensitive to be leaked?
    \end{itemize}
\end{itemize}

\emph{[Use ChatGPT plugin]:}
\begin{itemize}
    \item Have you used any ChatGPT plugins? What plugins did you use? What did you use them for?
    \item What data do you think the third-party plug-in developers can access? How do you think the data is used by the third party developers?
\end{itemize}

\subsection{Back to Reflections: Privacy Concerns and Challenges in Preserving Privacy}

Cool, we have discussed a lot about how ChatGPT (and other agents they use) works. 

\begin{itemize}
    \item Anything new or surprising for you today? 
    \item Given what you learned today, are there any additional concerns about sharing data with ChatGPT that you want to discuss with us?
    \item What do you think you can do to address these concerns?
    \item Do you wish for any improvement of existing features or tool support? 
    \item Do you think there should be any additional support that is not yet available?
    \item Let's imagine together. If you have a magic wand to change anything, what would the ideal support or scenario look like for you, that can make you feel totally secure about your data when using ChatGPT?

\end{itemize}

That's a wrap for our interview today! Thank you so much for taking the time to share your insights and experiences with us - it's been incredibly valuable. 
We hope you find it useful. Please don't hesitate to reach out if you have any further questions or thoughts you'd like to share. Thanks again for your contribution to our study!

\section{Codebook for the interview results}
\label{sec:codebook-interview}
Codebook for the interview results is shown in \autoref{tab:Codebook-interview}.

\begin{table*}[]
    \centering
    \caption{Codebook for the interview results}
    \begin{tabular}{p{0.1\linewidth} p{0.22\linewidth} p{0.6\linewidth}}
    \toprule
    Theme& Code& Memo\\
    \midrule
    Use Cases and Disclosure Behaviors& Ad Writing& Users shared details of items they were selling and their expectations to generate an ad.\\
    & Book Chapter Writing& Users provided topics or related contexts for drafting book chapters.\\
    & Career Advice& Users asked ChatGPT for career advice including generating templates or drafting resume or cover letter, or consulting, based on their educational and work experience and expectation.\\
    & Casual Chat& Users had casual chat with ChatGPT normally with less intention and just for fun or exploration. Many personal life information may be shared in casual chat.\\
    & Class Preparation& Users used ChatGPT for class preparation including brainstorming class sections or methods to teach certain things, normally shared the information related to class.\\
    & Concepts Learning& User used ChatGPT to help learning new concepts.\\
    & Copy-editing& Used ChatGPT for copy-writing, normally for work.\\
    & Create Apps& Users used ChatGPT to guide APP creation, including generating codes, debugging and problem solving.\\
    & Data Analysis& Users used LLM-based CAs for data analysis in many ways. From directly sharing raw data asking for results, explaining the tasks asking for solutions and just asked certain questions.\\
    & Diet Advice& User shared personal health related information or expectations to generate diet plan or ask for diet advice.\\
    & Email Writing& Users used ChatGPT for email writing in many ways. From directly sharing emails sent by others and asking for responses, to elaborating the context, drafting emails by users themselves and ask for revising or just provide general information to generate emails.\\
    & Email/Work Message Writing& Users used ChatGPT for work-related message or email writing or revising.\\
    & Exercise Advice& Users normally shared personal health related information and exercise goals for exercise advice.\\
    & Finance Advice& Users used ChatGPT for financial consultation such as debt problems, budget plan.\\
    & Generate Survey Responses& Users used ChatGPT to generate responses to the survey they have to do.\\
    & Immigration Advice& Users used ChatGPT to consult immigration problems such as visa application.\\
    & Info search& Users used ChatGPT for normal information search like Google search.\\
    & Joke Writing& Users used ChatGPT to generate jokes. Sometime provided their friends' names to generate ``personalized'' jokes.\\
    & Language Learning& Users used ChatGPT for language learning such as learning the expression through chatting.\\
    & Legal Advice& Users used agents for consulting legal related problems such as drafting contract, consulting related law based on given context.\\
    & Life Advice& Asking for life advice based on given personal conditions such as advice on work-life-balance.\\
    & Literature Search& Users used ChatGPT to search related literature based on certain topics or given content.\\
    & Marketing Advice& Asking for business marketing advice based on business conditions or general questions.\\
    & Math Learning& Used ChatGPT to learn math and prepare math exam.\\
    & Medical Advice& Asked for medical advice normally based on detailed personal medical or health information such as diagnosis results.\\
    & Portfolio Making& Used ChatGPT for brainstorming portfolio idea and for suggestions.\\
    & Programming& Used ChatGPT for programming for such as prototyping or debugging.\\
    & Relocation Advice& Used ChatGPT for suggestions about relocation. Users normally provided the location information to ask for suggestions.\\
    & Research Work& Users used ChatGPT to do research such as on clients (companies or persons) or certain topics.\\
    \bottomrule
    \end{tabular}
    \label{tab:Codebook-interview}
\end{table*}

\begin{table*}[]
    \centering
    \caption*{\autoref{tab:Codebook-interview} (continued)}
    \begin{tabular}{p{0.15\linewidth} p{0.25\linewidth} p{0.5\linewidth}}
    \toprule
    Theme& Code& Memo\\
    \midrule
    & Revise Writing& Users used ChatGPT to revise their own writing in various purpose such as paper revision.\\
    & Schoolwork& Used ChatGPT to help with or finish the schoolwork.\\
    & Social Media Post Writing& Used ChatGPT to generate social media post writing for personal, business or work purposes.\\
    & Test Capabilities& Users tries to test ChatGPT's capacities and explored what the agent can do.\\
    & Therapy& Users used ChatGPT as therapy to share personal life conditions, thoughts, feelings and emotion to more ask for emotional support.\\
    Factors that Affect Users’ Disclosure Intentions with LLM-Based CAs& Perceived Capability of the CAs& Users decide which agents to share or what information to share primarily considering the capability of the CAs they perceived. They tend to share more information if they believe the agents can help with their tasks while withdraw if they don't believe or unsatisfied with agents' capability.\\
    & Convenience of Operation& Users' disclosure intention or behavior can be influenced the convenience of operation. They may choose to share more or less depending on the operation convenience.\\
    & Perceived Personal Data Sensitivity& Users' disclosure would be influenced by the perceived sensitivity of the data. While the perceived sensitivity is vary from individuals.\\
    & Resignation& Users feel less concern to share information that can accessed from other places such as other online platforms or other databases.\\
    & Perceived Risks and Harms-Concerns over data misuse by institutions& Have concerns to share information because users unsure how their data will be used or whether will be misused by the companies behind the systems.\\
    & Perceived Risks and Harms-Concerns about others finding out& Users concerned on their usage of AI being discovered by others because unsure others' acceptance on using AI for certain tasks.\\
    & Perceived Risks and Harms-Concerns about idea theft& Users concerned on their ides such as business ideas, story ideas, or any unpublished works being theft by the people or companies behind the agents.\\
    & Perceived Norms of Disclosure-Attitude towards disclosing others' data and have one's own data shared by others& Users hold different attitudes on sharing others' data and have their own data shared by others have different intentions to disclosure (others') personal information.\\
    & Perceived Norms of Disclosure-Caution on sharing work-related data due to company policies or NDAs& Users normally are more cautious about sharing data from work because of the companies' policy or NDAs.\\
    How Users Navigate the Trade-off Between Disclosure Risks and Benefits& Accept Privacy Risks to Reap Benefits& For the benefits, some users choose to accept the privacy risks.\\
    & Avoid Certain Tasks Completely Due to Privacy Concerns& Some users just avoid using LLM-based CAs for certain tasks because they don't want to take the privacy risks on sharing certain information.\\
    & Manually Sanitize Inputs-Censor and/or Falsify Sensitive Information& Some users choose to censor or falsify sensitive information when sharing with the agents for protecting their privacy as well as reaping benefits.\\
    & Manually Sanitize Inputs-Desensitize Input Copied from Other Contexts& Some users desensitize the input copied from other context for protecting their privacy as well as reaping benefits.\\
    & Manually Sanitize Inputs-Only Seek General Advice& Some people choose to only ask for general advice instead of personalized ones for avoid sharing personal or detailed information.\\
    \bottomrule
    \end{tabular}
\end{table*}

\begin{table*}[]
    \centering
    \caption*{\autoref{tab:Codebook-interview} (continued)}
    \begin{tabular}{p{0.15\linewidth} p{0.22\linewidth} p{0.55\linewidth}}
    \toprule
    Theme& Code& Memo\\
    \midrule
    Mental Models of How LLM-Based CAs Handle User Input& Response Generation-A: ChatGPT is magic & Users with Mental Model A have limited understanding on how the system generates responses and they regard the system as an AI ``blackbox'' or ``magic'' thing.\\
    & Response Generation-B: ChatGPT is a super searcher & Users with Mental Model B think the responses generation process is seperated into different stages and components. AI is trained for certain stages such as search information from internet or databases, or for synthesizing results.\\
    & Response Generation-C: ChatGPT is a stochastic parrot & Users with Mental Model C have more comprehensive understanding on how the system generates responses. They know the system is based on an end-to-end ML model.\\
    & Improvement and Training-D: User input as a quality indicator& Users with Mental Model D think their input works as the quality indicator of the generated responses. They thought the users input used to train is to rate or give feedback so that decide what types of responses generated by the agents can be reused in the future.\\
    & Improvement and Training-E: User input is training data & Users with Mental Model E understand their input would be included into the training dataset. As the result, their input might be integrated into future responses. (While several people with this model thought they have their own personalized models)\\
    Users' Awareness and Reactions to Memorization Risks in LLM-based CAs& Having Encountered LLM Memorization Risks& Users said they have encountered memorization risks in their previous experience on other LLM-based application.\\
    & Having Concerns About Leaking Original Writing& Some users spontaneously shown concerns on their original writing shared being memorized and leaked to others, which may harm to novelty or authorship of their writing.\\
    & Unconcerned About Memorization Because Not Sharing Sensitive Data& Some users showed no concern on the memorization risks because they don't think their sharing included any sensitive data.\\
    & Unconcerned About Memorization Because They Could Not Imagine the Problem& Some users showed no more concern on the memorization risks because they cannot understand and see how the memorization happens.\\
    & Unconcerned About Memorization Due to the Perceived Deniability of the AI-Generated Output& Some users showed no more concern on data being memorized because they thought if the AI generate information about them that were inaccurate, then these data make no sense for them.\\
    Users' Awareness, Understanding, and Attitudes About Privacy-Protective Support& Users Lack Awareness of Existing Privacy-Protective Support& Users never thought about there would be some privacy-protective methods such as opt-out or using OpenAI API playground that they can use to protect their privacy meanwhile reap benefits from AI.\\
    & Dark Patterns Impeded Adoption of the Opt-out Control& There's dark patterns related to Opt-out control which restrains users' usage on the Opt-out to do data control.\\
    & Users Anticipated More Granular Opt-out Control& Users hoped they can have more granular on the opt-out control\\
    \bottomrule
    \end{tabular}
\end{table*}

\end{document}